# Equatorial scattering and the structure of the broad-line region in Seyfert nuclei: evidence for a rotating disc

J. E. Smith[1]*, A. Robinson[1 †], S. Young[1], D. J. Axon[1 †], Elizabeth A. Corbett[2]

1 Centre for Astrophysics Research, University of Hertfordshire, Hatfield, Hertfordshire, AL10 9AB, UK
2 Anglo-Australian Observatory, PO Box 296, Epping, NSW 1710, Australia

**ABSTRACT**

We present detailed scattering models confirming that distinctive variations in polarization across the broad H$\alpha$ line, which are observed in a significant fraction of Seyfert 1 galaxies, can be understood in terms of a rotating line-emitting disc surrounded by a co-planar scattering region (the equatorial scattering region). The predicted polarization properties are: averaged over wavelength, the position angle of polarization is aligned with the projected disc rotation axis and hence also with the radio source axis; (ii) the polarization PA rotates across the line profile, reaching equal but opposite (relative to the continuum PA) rotations in the blue and red wings; (iii) the degree of polarization peaks in the line wings and passes through a minimum in the line core. We identify 11 objects which exhibit these features to different degrees. In order to reproduce the large amplitude PA rotations observed in some cases, the scattering region must closely surround the emission disc and the latter must itself be a relatively narrow annulus — presumably the H$\alpha$-emitting zone of a larger accretion disc. Asymmetries in the polarization spectra may be attributable to several possible causes, including bulk radial infall in the equatorial scattering region, or contamination by polar scattered light. The broad H$\alpha$ lines do not, in general, exhibit double-peaked profiles, suggesting that a second H$\alpha$-emitting component of the broad-line region is present, in addition to the disc.

**Key words:** polarization – scattering – galaxies: active – galaxies: Seyfert

## 1 INTRODUCTION

Although the Seyfert Unification Scheme (e.g. Antonucci 1993) is now widely accepted, important questions remain concerning the geometry, structure and kinematics of Seyfert nuclei – particularly on sub-torus scales. A major obstacle here is that the inner structures of active galactic nuclei (e.g. the continuum source, accretion flow and broad-line region; BLR) are too compact to be studied directly by the currently available observational techniques. A greater understanding of the BLR is especially desirable, not merely because the broad emission lines are among the most prominent features in AGN spectra, but also because they are currently the only means of probing the massive black hole mass function over a large range in redshift and hence also luminosity (e.g. Kaspi et al. 2000; Laor 2001; McLure & Dunlop 2001, 2002).

---

* E-mail: jsmith@star.herts.ac.uk
† Current address: Department of Physics, Rochester Institute of Technology, 85 Lomb Memorial Drive, Rochester, NY 14623-5603, USA

Indirect techniques such as reverberation mapping (e.g. Peterson 1993) can, in principle, provide information concerning the spatial distribution of the emitting gas, but in practice require long monitoring campaigns and are often hindered by technical and interpretational difficulties (e.g. Peterson, Pogge & Wanders 1999). Whilst these studies appear to favour Keplerian dynamics, it is unclear whether or not the gas distribution has axial or spherical symmetry (e.g. Wanders et al. 1995; Peterson & Wandel 2000).

Spectropolarimetry provides an alternative approach to investigating the nature of the BLR. The unique diagnostic strength of this technique is that the polarization state of scattered light carries the imprint of the scattering geometry, allowing the structure and kinematics of both the scattering medium and the emission source to be investigated in unresolved sources. In particular, spectropolarimetry of Seyfert 1 galaxies and broad-line radio galaxies often reveals striking variations in both the degree ($p$) and position angle of polarization ($\theta$) as a function of wavelength across the broad H$\alpha$ emission line (e.g. Goodrich & Miller 1994, hereafter GM94; Martel 1996; Corbett et al. 1998, 2000; Cohen & Martel 2001, hereafter CM01; Smith et al. 2002, hereafter S02), which may provide important clues to the structure and kinematics of the BLR and the associated scattering regions.

A key constraint on the scattering geometry is the orientation of the polarization position angle relative to the projected radio source axis, assuming that the latter traces the symmetry axis of the nucleus. In most Seyfert 2 galaxies and a minority of Seyfert 1 galaxies, the optical polarization position angle (PA) is perpendicular to the projected radio source axis (e.g. Antonucci 1983; Brindle et al. 1990) – as expected if polar scattering dominates the observed polarization. Typically, however, the optical polarization properties of Seyfert 1 galaxies are *not* consistent with polar scattering. In Seyfert 1 galaxies the optical polarization PA is more often *parallel* to the radio axis Smith et al. (2004, and references therein). At least some of the scattered light emerging from the nucleus must therefore follow a different path to that in the Seyfert 2 galaxies. This in turn suggests that an additional source of polarized light is present in Seyfert nuclei, which dominates in most Seyfert 1's but is not seen in Seyfert 2's.

In S02 we proposed that the optical polarization properties of Seyfert nuclei can be explained by a model in which the broad Balmer line emission originates in a rotating disc and is scattered by two distinct scattering regions that produce orthogonally polarized light. These are:

(i)   a compact scattering region co-planar with the line-emitting disc and situated within the torus, in its equatorial plane – the 'equatorial' scattering region;

(ii)  a scattering cone situated outside the torus but aligned with the torus/emission disc axis – the well established 'polar' scattering region.

As outlined in S02 and discussed in more detail in Smith et al. (2004; hereafter S04) much of the observed range in polarization properties exhibited by Seyfert nuclei can be broadly understood in terms of an orientation sequence based on this two-component scattering model. In Seyfert 2 galaxies, the equatorial scattering region is hidden inside the torus, and therefore the observed polarization is dominated by polar scattering. In some cases, the scattered flux produced is sufficient for broad lines to be detected in polarized light (e.g. NGC 1068; Antonucci & Miller 1985). A significant minority (10-30 per cent) of Seyfert 1 galaxies also appear to be dominated by polar scattering. We argue (S04) that they are viewed through the upper layers of the torus and are thus subject to moderate extinction, sufficient to suppress polarized light from the equatorial scattering region, but not the broad wings of the Balmer lines. In most other Seyfert 1 galaxies, the equatorial region dominates, producing polarization parallel to the projected radio source axis. Equatorial scattering of light from a rotating disc also explains the characteristic polarization structures that are observed across the broad H$\alpha$ lines in many Seyfert 1's (Smith 2002; S02), including large position angle rotations and distinctive variations in percentage polarization.

In S02 we presented optical polarization spectra of a sample of thirty-six Seyfert 1 galaxies. In S04 we discussed those Seyfert 1 galaxies whose polarization characteristics are best explained by 'polar' scattering. We also outlined our orientation dependent model for the polarization properties of Seyfert nuclei. In this paper, we investigate in more detail the polarization signatures produced by the equatorial scattering region and discuss the Seyfert 1 galaxies in which we believe that this scattering geometry dominates. In Section 2 we introduce the basic properties of our equatorial scattering model and in Section 3 explore how varying the parameters of the basic model influences the resultant polarization spectrum. In Section 4 we discuss the net polarization spectrum resulting from the combination of equatorial and polar scattering. In Section 5 we list the Seyfert 1's whose broad H$\alpha$ polarization we believe to be dominated by equatorial scattering and briefly discuss their properties in relation to the models. Finally, we explore the implications of our equatorial scattering model for the structure and kinematics of the BLR and speculate on the nature of the equatorial scattering region.

## 2 EQUATORIAL SCATTERING

Our spectropolarimetric study of Seyfert 1's (S02) revealed that the polarization structure across the broad H$\alpha$ line varies widely from object to object. Nevertheless, we identified three broad categories, which can be associated with different orientation regimes of our two component scattering model (S04). In this scheme, 'null polarization' objects (those with low measured polarizations) are viewed face on, i.e., the line-of-sight is close to the symmetry axis of the active nucleus. Objects exhibiting polarization signatures of polar scattering are those viewed at grazing incidence to the torus. At intermediate viewing angles, equatorial scattering dominates the observed polarization. In this paper, we will focus on the latter category. While their polarization properties differ in detail from object to object, it is possible to discern certain common characteristics that are present to differing degrees in most objects. These are (i) a swing in position angle across the H$\alpha$ profile and (ii) a dip in polarization in the core of the profile, flanked by polarization peaks in the wings.

Within the group of equatorially scattered objects, the wavelength averaged polarization PA tends to be aligned with the projected radio source axis in cases where the latter can be reliably determined. This indicates that the scattering takes place in a plane perpendicular to the radio axis. A plausible scenario is that the emission source is surrounded by an optically thin scattering disc that is co-axial with the radio source (Antonucci 1984; GM94). In the context of the Seyfert unification scheme the scattering disc must lie in or close to the equatorial plane of the torus.

As has previously been discussed by several authors (GM94; Cohen et al. 1999; CM01), PA swings across the broad lines can be produced if the source of the line emission is a rotating disc and light from this disc is scattered in a co-planar scattering region. However, PA swings over broad H$\alpha$ are often accompanied by the variations in $p$ outlined above. We argue in Smith (2002) and S02 that the combination of a rotating emission disc and this 'equatorial' scattering geometry can simultaneously explain not only the variations in $\theta$ but also the characteristic variations in $p$.

The main features of the equatorial scattering geometry outlined in S02 are sketched in Fig. 1. A rotating line-emitting disc is centred on a point-like continuum source, $C$, and is tilted at an arbitrary inclination to the line-of-sight. The disc is surrounded by a co-planar ring of scatterers, which, in turn, is surrounded by the circum-nuclear torus. The disc and ring are co-axial with the torus and lie in its equatorial plane. We assume that the radio source is also aligned with this axis, which is, therefore, the symmetry axis of the system.

Light from the emission disc and from the central continuum source is scattered from each point in the scattering ring into the observer's line-of-sight and thereby becomes linearly polarized. For points (hereafter, scattering elements) aligned with the major axis of the emission disc (as seen in projection), the **E** vector of the scattered continuum light will be parallel, in projection, to the disc axis. The scattering element will 'see' red- and blue-shifted line emission from locations $A$ and $B$ in the disc, respectively. Since the scattering planes for red- and blue-shifted rays are different, the **E** vectors of the corresponding scattered, polarized rays are directed either side of the continuum **E** vector with equal, but opposite offset angles. This leads to a rotation in polarization PA ($\theta$) from blue-to-red across the broad-line profile, centred on the continuum PA.

The model also predicts distinctive variations in $p(\lambda)$ across the broad-line profile. All scattering elements around the ring have an edge-on view of the disc, and thus 'see' a broader emission-line profile than the direct line-of-sight, which for a Seyfert 1 nucleus, must pass within the 'funnel' of the torus. As a result, the emission-line profile in direct, unpolarized, light is narrower than that in scattered, polarized, light (see Fig. 1) and the combination of scattered and direct line emission results in wavelength-dependent dilution of the polarized component. With the addition of the underlying continuum, the observed percentage polarization increases relative to the continuum polarization in the line wings, but drops to a minimum below that level in the line core.

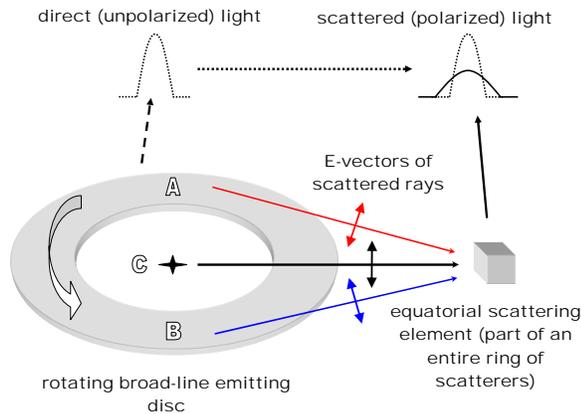

**Fig. 1.** The equatorial scattering geometry. The central continuum source (C) and a broad-line emitting disc are surrounded by a co-planar ring of scattering particles (only a single element of the scattering ring is shown). Scattering of light from the central continuum source and the broad-line emitting disc in the plane of the disc produces characteristic variation in both $p$ and $\theta$ across the broad-line profile (see text).

## 2.1 The Generic Scattering Model

The polarization signature of the equatorial scattering model can be understood in general terms using the conceptual arguments described above. However, in order to calculate $p(\lambda)$ and $\theta(\lambda)$ in detail and to explore the effects of varying key parameters, we have employed the computer code that we used for our detailed study of Mrk 509 (Young et al. 1999). Earlier versions of this code (hereafter referred to as the Generic Scattering Model or GSM) have been used in several previous polarization studies of AGN (e.g. Young 1995; Young et al. 1995; Packham et al. 1997; Young et al. 1998). A detailed description of the version used here can be found in Young (2000).

For specified configurations of the emission source and scattering geometries, the code computes the integrated Stokes parameters of the scattered light. Both the source and scattering regions are modelled as a number of discrete elements, which can be arranged in a variety of different configurations. In the calculations presented here, we consider only the case where the emission source is a rotating disc. The disc is geometrically thin and divided up into a number of source elements in both the radial and azimuthal directions, each of which produces a Gaussian emission-line profile. The FWHM of the elemental line profile represents the velocity dispersion of the emitting gas within the element. The number of elements in each direction can be adjusted as appropriate for the geometry being investigated, subject to adequate sampling of the velocity field of the disc and the scattering angles it subtends at the scattering medium. The rotational velocity and line emissivity distribution are specified as radial power laws, but in the models considered here, the velocity field is always Keplerian and the surface emissivity is constant. The continuum emission is modelled as originating in the emission disc itself (Young 2000). However, in practice the continuum polarization produced in the models presented here behaves as though the emission source is at the centre of the system, since the disc has azimuthal symmetry and the continuum flux density is wavelength independent.

The scattering regions are likewise divided into discrete elements. Both the polar scattering region and a compact equatorial scattering region are modelled. The polar scattering region is represented as a radially truncated cone whilst the equatorial scattering region is modelled as a flared disc. These regions are separately specified by half-opening angles and inner and outer radii. In the models considered here, both structures have their symmetry axes aligned with the rotation axis of the emission disc, which we also take to be the principal symmetry axis of the nucleus. The scattering particles in both regions are assumed to be free electrons whose number density has a radial dependence given by a power law. Each scattering region is also characterised by a kinetic temperature and a bulk velocity field (e.g. rotation or outflow).

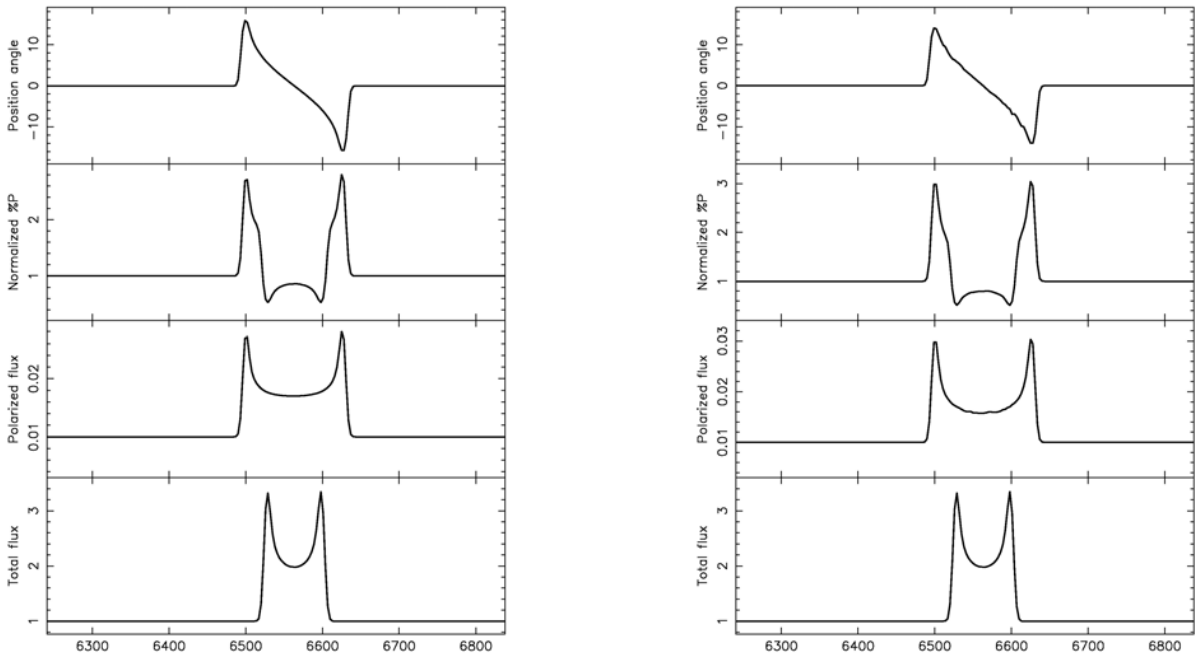

**Figure 2.** (**a**, left panel) Polarization spectra for an isolated, stationary clump of scatterers located in the plane of a rotating emission ring. The panels show, from the bottom, the total flux, the polarized flux, the percentage polarization, and the position angle of polarization. The percentage polarization is normalized such that the continuum polarization has a value $p = 1$ per cent. (**b**, right panel) as Fig. 2(a) for a stationary scattering ring co-planar with a rotating emission ring. In both cases, the system inclination $i = 35°$ and the emission ring has a rotation speed $v_r = 3000$ km s$^{-1}$.

The GSM implicitly assumes that the scattering regions are optically thin, so that only single scattering need be considered. This is probably a good approximation for the polar scattering region but may be less secure for the equatorial scattering region, whose properties (apart from its basic geometry) are unknown. Nevertheless, simulations by Henney & Axon (1995) suggest that even for optically thick scattering media, the polarization is dominated by first order scattering and therefore, the resultant polarization signature approximates to that for single scattering.

### 2.2 Scattering from clumps and rings

We first consider the simplest configurations for the emission source and scattering region consistent with the equatorial scattering model. The BLR is represented as a thin rotating ring and we discuss cases in which the scattering region is, respectively, an isolated clump and a thin, stationary ring. The polarization spectra produced by these configurations are relatively easy to interpret and provide insights into some important effects that are more difficult to isolate in the polarization spectra produced by geometrically extended emission and scattering regions. A particular consequence of modelling the BLR as a rotating ring is that both the directly viewed and scattered broad-line profiles will, in general, have double-peaked structures. However, the general form of the variations in $p(\lambda)$ and $\theta$ predicted by the conceptual reasoning outlined above remain clear. In the models discussed below, both the emission (BLR) and the scattering rings have a nominal radial thickness of 1-m.

#### 2.2.1 Scattering from an isolated clump

The polarization spectrum produced by scattering of line emission from a rotating thin ring by an isolated 'clump' of scatterers is shown in Fig. 2(a). The scattering clump was modelled within the GSM by invoking the polar scattering cone, with parameters chosen such that its symmetry axis lies along the projected major axis of the emission ring. The line-emitting ring was composed of 180 source elements in the azimuthal direction, each of which produces a Gaussian line of Full Width Half Maximum (FWHM) 380 km s$^{-1}$. It is inclined at an angle of 35° relative to the line-of-sight.

As the scattering clump is located on the projected major axis of the emission ring, the continuum has a polarization PA of 0°, orthogonal to this axis and thus aligned with the symmetry axis of the system. Within the emission line, the polarization PA exhibits equal but opposite rotations (relative to the continuum) in the blue and red wings. There is a sharp swing to a positive offset in the blue wing. The direction of the swing then reverses through the line centre to reach a negative offset of equal magnitude in the red wing. The sense of the PA rotation would be reversed (i.e. with the positive offset in the red wing) if the emission ring were rotating in the opposite direction. From the point of view of the scattering clump, line emission originating at those points where the line-of-sight is tangential to the ring has the greatest doppler-shift from the rest wavelength of the line. Scattered light from these points also has the greatest PA offset from the continuum value. As a consequence, the greatest deviation (~16°) from the continuum PA (0°) occurs at the extremities of polarized broad-line wings. The PA is 0° at the line core (i.e. the same as the continuum) since the polarized emission here originates in source elements located along the line-of-sight to the centre of the system, for which the doppler-shift, as seen by the scattering clump, is also zero.

Significant variations in $p(\lambda)$ also occur across the broad-line profile, with $p(\lambda)$ dropping below the continuum value in the line core but peaking above the continuum level in the extreme wings. As discussed above, this structure is due to differential dilution of the scattered line by direct emission, which has a narrower line profile. In this particular model, $p(\lambda)$ exhibits relatively complex structure since both direct and scattered profiles are double-peaked, resulting in two separate polarization minima on either side of the line core.

### 2.2.2   *Scattering from a narrow ring*

Very similar variations in $p(\lambda)$ and $\theta(\lambda)$ are produced if we replace the isolated clump with a complete scattering ring. The scattering ring is represented within the GSM, by invoking the compact scattering region with appropriately small values for the outer radius and half-opening angle. This structure is co-planar with, and completely surrounds, the line-emitting ring, which is itself tilted at an inclination of 35° to the line-of-sight. The polarization spectrum for this configuration is shown in Fig. 2(b). The variations in both $p(\lambda)$ and $\theta$ over the broad-line are essentially identical in form to those produced by the isolated clump of scatterers and shown in Fig. 2(a). At first sight, this seems counter-intuitive, since for a circular scattering ring we might expect symmetry to dictate complete cancellation of the Stokes parameters from points around the ring, resulting in zero net polarization. Complete cancellation will indeed occur if the system is viewed face-on ($i=0°$), but if the ring is inclined with respect to the line-of-sight, differences in scattering angle around the ring ensure that there is a net polarization dominated by light scattered from the elements located along the projected major axis of the emission ring (i.e., where this axis intersects the scattering ring).

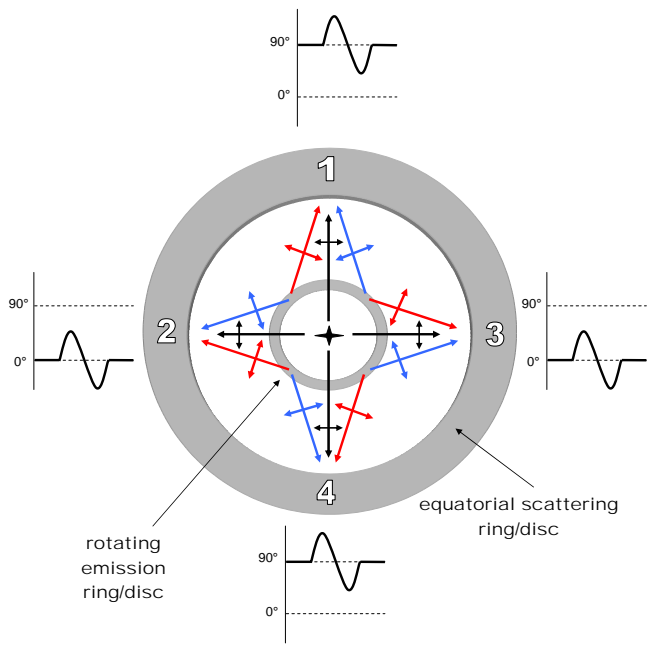

**Figure 3.** Polarization PA rotations across the scattered line profiles at cardinal points of the scattering ring. The PA rotation has the same form across the line profile at each point but the average PA is parallel to the axis at points 2 and 3 and perpendicular to it at points 1 and 4.

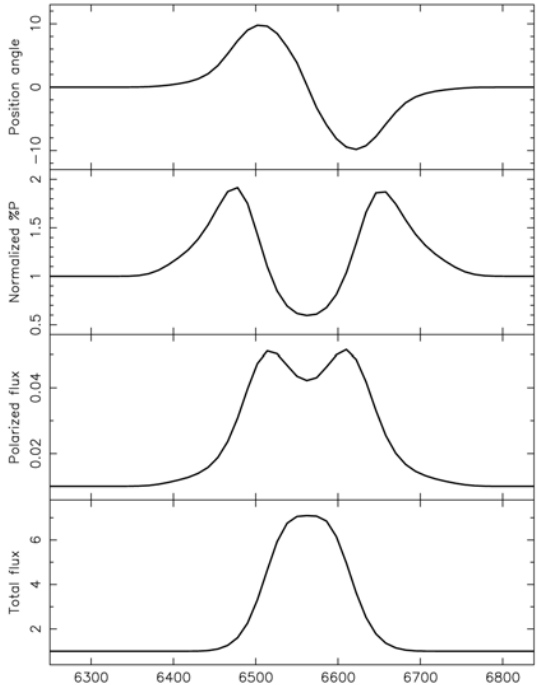

**Figure 4.** As Fig. 2(a) for the Standard Equatorial Model: a rotating emission disc surrounded by a stationary, co-planar 'disc' of scatterers.

This effect merits more detailed explanation. We first consider scattering elements located along the projected major and minor axes of the inclined disc (labelled 1–4 in Fig. 3). Given that the compact scattering region is stationary and confined to the equatorial plane, each scattering element produces a line profile that is identical in shape and central wavelength. The polarization PA varies across each of these elemental scattered line profiles just as for an isolated clump of scatterers [Fig. 2(a)]. However, whilst the *form* of the PA rotation is identical the *average* PA (or equivalently, the continuum PA) differs around the scattering ring. The polarized continuum has a PA perpendicular to the projection of the line joining the scattering element to the centre of the system and the rotations over the red and blue wings of the emission line are symmetric about the continuum PA (Fig. 3).

When $i=0$, the net polarization (after performing an azimuthal intergration around the ring) is zero since there is complete cancellation between orthogonally polarized light from perpendicular scattering elements (e.g. points 1 and 2 in Fig. 3). However, as $i$ increases, the scattering angle decreases for most elements around the ring with the result that light scattered from these elements is less completely polarized. Conversely, for elements located on the projected major axis of the line emitting disc (points 2 and 3 in Fig. 3), the scattering angle remains unchanged, at $\sim 90°$, These elements therefore produce a greater *polarized flux* than their orthogonal counterparts (i.e. points 1 and 4). This breaks the symmetry and leaves a net polarization with a continuum PA and broad-line PA rotation similar to that produced by a single scattering element aligned with the projected major axis of the emission disc.

For intermediate points around the scattering ring, the scattering planes of the blue- and red-shifted rays from the emission disc are asymmetrically tilted, in general (for arbitrary values of $i$), with respect to that of the continuum ray. There is a similar asymmetry in the scattering angles. This results in a corresponding asymmetry in the scattered line profile from each element, the red and blue peaks having different $p$ and PA offsets relative to the continuum. However, these asymmetries vanish upon azimuthal integration around the complete scattering ring because each intermediate scattering element has a 'complementary' element[1] that produces a mirror polarized line profile whose asymmetry is in the opposite sense. It follows that if part of the scattering ring is either missing, or obscured, asymmetries will be induced in $p$ and $\theta$ between the red and blue sides of the integrated scattered line profile.

CM01 have previously discussed a model in which PA rotations across the broad-line profile arise from a partially obscured scattering ring around a rotating emission ring. Obscuration is invoked to provide the asymmetry required to produce the PA rotation. The foregoing discussion shows that this is not, in fact, necessary. A PA rotation arises as a natural consequence of the scatterers having different lines-of-sight to red and blueshifted emission from opposite sides of the emission disc, coupled with the dependence of $p$ on scattering angle. The latter, in effect, acts as the 'symmetry breaker' and results in light scattered from along the major axes of the projected emission disc dominating the resultant polarization. We return to the issue of incomplete scattering rings and asymmetric

---

[1] Located at the supplementary angle relative to the major axis.

polarization of the red and blue line wings in Sections 5.1.1 and 5.1.2.

### 2.3 Radially extended emission and scattering regions

Whilst it is conceptually useful to consider concentric 'rings' of line-emitting gas and scattering material, a radially extended disc or annulus is almost certainly a more accurate representation of both the broad-line region and the scattering regions. It has been suggested that the BLR, for example, covers a factor of at least 10 in radius (e.g. Vestergaard, Wilkes and Barthel 2000) and it is likely that any scattering region is also radially extended. Our equatorial scattering model should therefore sample the radial, as well as the azimuthal direction, within the emitting and scattering regions.

As our aim is to investigate the polarization signature of scattering from the equatorial plane of an emission disc, rather than the detailed shapes of the broad-line profiles, or the nature of the emission disc itself, we adopt a simple model defined by a standard set of parameters. In all models discussed below, the region emitting the broad-lines is represented as a thin rotating disc, of uniform surface emissivity, whose outer radius $R_{outer} = 10 \times R_{inner}$, where $R_{inner}$ is the inner radius. The disc is composed of 256 source elements (16 in both radial and azimuthal directions). Each element emits an intrinsic Gaussian line profile of FWHM 2280 km s$^{-1}$, centred at the local Keplerian rotation velocity. This value for the intrinsic profile FWHM is chosen in order to produce a smooth, single-peaked broad-line profile in total flux, rather than the double-peaked profile that the disc would produce in the absence of local broadening (double-peaked profiles are rarely observed in Seyfert 1 galaxies, Section 5.3).

In Fig. 4, we present the polarization spectrum for our base model, consisting of the emission disc described above surrounded by a radially extended equatorial scattering region. The scattering region is modelled as a thin circular wedge with a fractional width (relative to the inner radius) of 0.6. We henceforth refer to this model, (for which, parameter values are listed in Table 1) as the 'Standard Equatorial Model' (SEM). In the case shown in Fig. 4, the system axis is tilted at the same inclination ($i$=35°) as for the concentric rings model discussed in Section 2.2 [Fig. 2(b)].

**Table 1.** Parameters for Standard Equatorial Model.

| | |
|---|---|
| system inclination | 35° |
| emission disc inner radius | $1.0\times10^{14}$ m |
| emission disc outer radius | $1.0\times10^{15}$ m |
| rotational velocity at inner radius of emission disc | 7500 km s$^{-1}$ |
| compact scattering region half opening angle | 1° |
| compact scattering region inner radius | $1.5\times10^{15}$ m |
| compact scattering region outer radius | $2.4\times10^{15}$ m |
| electron number density at inner radius of compact scattering region | $1.0\times10^{12}$ m$^{-3}$ |
| number density radial dependence in compact scattering region | r$^{-1}$ |

The $p$ and $\theta$ spectra exhibit the same basic features as those of the concentric rings model. The polarization PA exhibits equal but opposite rotations relative to the continuum in the blue and red wings of the line and the percentage polarization has a central minimum in the core of the line, flanked by peaks in either wing.

The most obvious difference is that the SEM produces much smoother spectra than the concentric ring model. This is partly due to the larger local broadening, but subtler effects are also important.

As explained in Section 2.2.1, the points on the line-emitting ring that produce light with the largest Doppler-shifts are also those that subtend the largest angles at any point on the scattering ring. In the polarized line profile, therefore, the largest PA offsets relative to the continuum occur at the maximum velocity shift from line centre. In contrast, for a radially extended emission disc with a Keplerian velocity field, the source elements emitting light with the greatest Doppler-shift from the line core, as seen by any scattering element, are located at the inner edge of the emission disc. These source elements also subtend the smallest angle at any point on the scattering ring and as a result, the far wings of the scattered polarized line profiles produced by each scattering element have the smallest PA offsets relative to the continuum (see Fig. 5). Dilution of the line flux by the underlying continuum ensures that the PA in the far wings of the line converges smoothly with the continuum PA. Conversely, source elements located around the outer edge of the emission disc at relatively large radii produce emission with both the smallest Doppler-shift and the largest PA offset. Therefore, the PA varies smoothly through the line profile, reaching equal but opposite maximum offsets relative to the continuum PA in each wing.

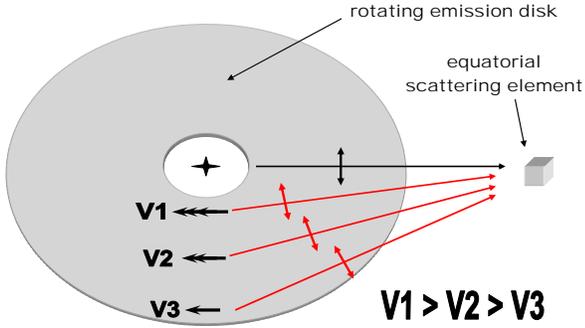

**Figure 5.** Polarization PA rotations for an extended emission disc. Line flux originating near the inner radius of the disc is polarized with a PA closer to that of the central continuum source than that originating at larger radii. For a Keplerian disc, the line flux originating near the inner radius also has the greatest Doppler-shift from the rest wavelength, as seen by the scatterers, and hence forms the wings of the line profile in polarized flux.

With the SEM parameters the polarized flux profile retains a double-peaked shape, as is characteristic for line emission originating in a disc viewed close to the plane of rotation – i.e., the scatterers' view. However, since the total flux profile has a broad central peak, the $p$ spectrum exhibits a single smooth minimum at the line core.

## 3  KEY PARAMETERS OF THE STANDARD EQUATORIAL MODEL

The basic equatorial scattering model, as so far outlined, can reproduce the general structure of the polarization variations commonly seen over the broad H$\alpha$ emission line in Seyfert 1 galaxies; namely the blue-to-red polarization PA rotation and the peak-trough-peak variation in $p$. There are, however, considerable differences in detail between the

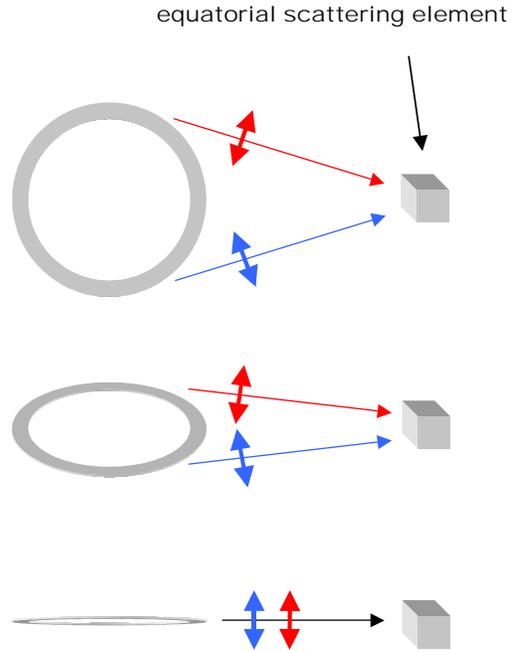

**Figure 6.** Effect of system inclination on PA rotation amplitude. The scattering planes for red and blueshifted emission from opposite sides of the rotating emission disc converge as the system inclination is increased.

polarization spectra $p(\lambda)$ and $\theta(\lambda)$ of different objects. It is therefore of interest to investigate how the variations in $p$ and $\theta$ across the line profile are affected by varying selected parameters of the equatorial scattering model

For this purpose we are only interested in those parameters that affect the form and amplitude of the variations in $p$ and $\theta$, rather than those that simply determine the level of polarization. The former are those parameters that significantly influence either the scattering geometry (system inclination, emission disc – scattering region distance) or the wavelength of the scattered ray (bulk velocity field of the scattering region). The latter include parameters, whose main effect is to modify the effective covering factor of the scattering region (e.g. the scattering region half-opening angle[2]) or the scattering column density (e.g. the scattering particle density distribution). In each of the cases described below, we start with the SEM and explore the effects of varying a single selected parameter.

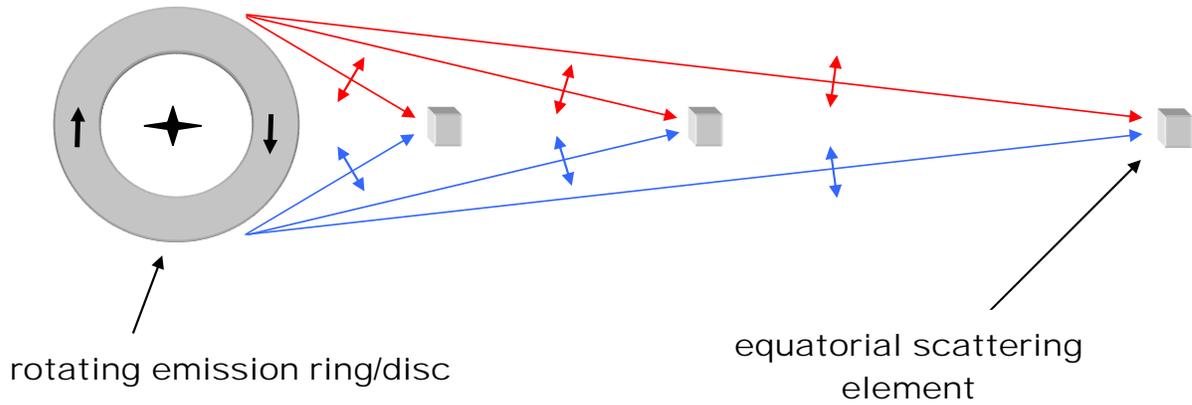

**Figure 7.** Effect of emission disc – scattering element distance on PA rotation amplitude. The scattering planes for red and blueshifted emission from opposite sides of the rotating emission disc converge as the distance between the line-emitting disc and the surrounding scattering ring is increased.

### 3.1  System inclination

As inclination is the key parameter in the Seyfert unification scheme and as the equatorial scattering geometry itself is an axisymmetric structure, it is clearly of interest to explore how the polarization properties vary with inclination.

The characteristic variations in $p$ and $\theta$ across the line profile are present for all inclinations $0<i<90°$. When $i=0°$, circular symmetry leads to complete cancellation of the **E** vectors and thus, null polarization. As the inclination is increased, the scattering planes for red- and blueshifted emission from opposite sides of the emission disc converge (Fig. 6) and the amplitude of the PA rotation decreases. The form of the variation in $p$ over the broad-line also changes since, as inclination increases, the line profile in direct light broadens, modifying the degree of dilution as a function of wavelength (in the limiting case, when $i=90°$, the directly viewed and scattered profiles are exactly the same, and $p$ is constant). However, the magnitude of both effects is relatively small for inclinations $i<45°$, the inclination range within which, according to the unification scheme, Seyfert 1 galaxies are found. The inclination-dependences of $p(\lambda)$ and $\theta(\lambda)$ are therefore unlikely to be a source of significant diversity in Seyfert 1 polarization spectra.

The *degree* of polarization is, however, rather more sensitive to inclination, even for $i<45°$. When $i\sim0°$, light scattered from points on the scattering ring located along the projected minor axes of the emission disc has a scattering angle $\sim90°$ and hence a relatively high polarization. As a result, nearly complete cancellation occurs with the orthogonally polarized light scattered from points along the projected major axes. Therefore, although the amplitude of the PA rotation is maximal at very small inclinations, the degree of polarization approaches zero. As the inclination increases, however, light scattered from points along the projected minor axis is less polarized, there is less cancellation with polarized light scattered from projected major axes (whose scattering angle remains at $\sim90°$) and hence the net polarization increases (see S04, fig. 8).

---

[2] Increasing the scattering region half-opening angle beyond $\sim45°$ would affect the form of $p(\lambda)$ and $\theta(\lambda)$, but such a large value would be inconsistent with our basic premise that the compact scattering region is closely confined to the equatorial plane of the circum-nuclear torus.

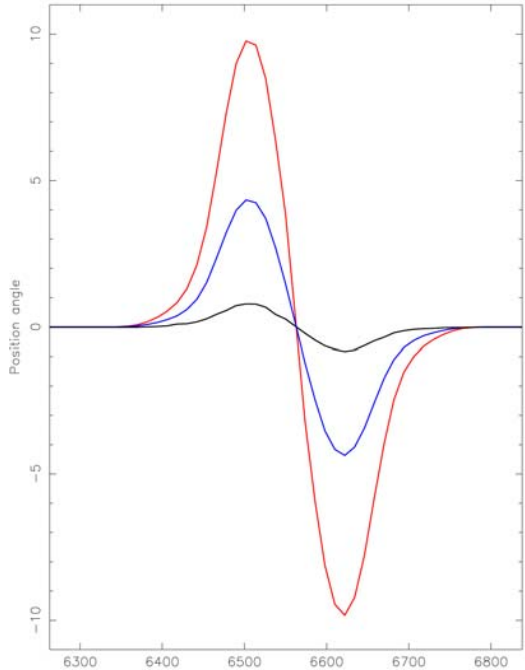

**Figure 8.** Polarization PA spectra for increasing values of the scattering region radius. For the models shown, the ratio of the inner radius of the scattering region to the inner radius of the emission disc is 15 (red line; the SEM), 40 (blue line) and 240 (black line). In each case the radial extent of the scattering region is $9.0 \times 10^{14}$ m (as in the SEM).

### 3.2 Emission disc–equatorial scattering region distance

The amplitude of the PA rotation across the broad-line profile also depends on the distance between the emission disc and the scattering material. As this distance increases, the scattering planes for red and blueshifted rays converge (Fig. 7) and the amplitude of the PA rotation is reduced. If the scattering material is removed to a sufficiently large radius, the scattering planes become well-enough aligned that the PA rotation vanishes. Models showing the effect of increasing the inner radius of the scattering region relative to the inner radius of the emission disc are presented in Fig. 8. Note that although the amplitude of the PA rotation is strongly dependent on this parameter, the form of $\theta(\lambda)$, and the spectral range over which the variation occurs, are unchanged.

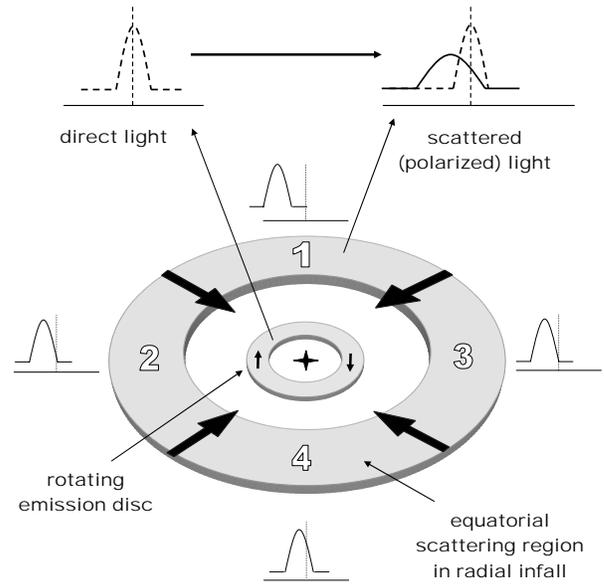

**Figure 9.** Equatorial scattering region undergoing radial infall. The doppler-shifts imparted to the scattered line profiles at different points around the region are shown schematically.

The form of $p(\lambda)$ is similarly unaffected, since the shapes of both the directly viewed and the scattered line profiles are independent of the scattering region radius. However, less light from the emission disc is intercepted by the scattering region as its radius is increased and thus the magnitude of $p$ decreases uniformly across the spectrum.

### 3.3 Bulk motions of the equatorial scattering material

In the models we have considered so far, the scattering region is assumed to be static. However, given the strong gravitational and radiation fields that are present in active nuclei, it is probable that the scattering medium is undergoing some form of bulk motion. The presence of radial flows in AGN is well established. For example in Broad Absorption Line Quasi-Stellar Objects (BALQSO's), blue-shifted absorption line systems indicate ionised material outflowing from the nucleus (e.g. Weymann et al. 1991). More pertinently, there is also evidence for scattering outflows along the poles of the torus in some Seyfert 2 galaxies, where the polarized broad-lines are often redshifted by several hundred km s$^{-1}$ (e.g. Miller, Goodrich & Matthews 1991; Young et al. 1993). A similar polar scattering outflow with a velocity of ~3000 km s$^{-1}$ has been found in the quasar 4C 74.26 (Robinson et al. 1999).

Evidence for other types of bulk motion is less obvious. However, both radial infall and rotational motions are plausible and indeed, there must be an inward gas flow in order to maintain a supply of fuel to the central engine. It is likely that this accretion flow will be concentrated in the equatorial plane of the torus and it will thus pass through the location of our postulated equatorial scattering region.

In this section, we consider the effect on the polarization structure across the broad-line profile if the equatorial scattering region is undergoing, respectively, radial or rotational bulk motions. In the cases we consider here, it is assumed that the motions are confined to the equatorial plane. All other parameters are the same as for the SEM (Table 1).

### 3.3.1   *Radial flows*

As an example of radial motion, we consider radial infall in the equatorial plane. For simplicity we assume that the radial velocity is uniform over the scattering region – i.e., all scattering elements have the same velocity, $v_s$. In general, the wavelength of the outgoing scattered ray is subject to 2 doppler shifts – one due to the relative motion between the source and scattering elements and the other due to the relative motion between the scattering element and the observer. Light emitted by the disc and incident on any element of the scattering region receives a blueshift relative to its emission wavelength, the magnitude of which depends (since the BLR is resolved at the scattering region) on the component of the scattering element velocity in the direction of the source element. Similarly, the scattered light receives a doppler-shift relative to its incident wavelength which depends on the velocity component in the direction to the observer. This, in turn depends on the azimuthal angle, $\alpha$, locating the element (measured from the intersection of the minor axis with the scattering ring – point 4 in Fig. 9) and the inclination *i*: $v_{los} = v_s \cos\alpha \sin i$. The doppler-shift of the scattered light relative to its incident wavelength varies from a redshift $+(v_s/c)\sin i$ (point 4 in Fig. 9) to a blueshift $-(v_s/c)\sin i$ (point 1 in Fig. 9). The scattered line profile produced by a given scattering element has a doppler-shift, relative to its emitted central wavelength, given by the product of the doppler-shifts imparted to the incident and scattered rays. The final doppler-shifts are shown schematically in Fig. 9 for the points at which the projected major and minor axes cross the scattering annulus. For this case (radial infall) all 4 elemental line profiles are blue-shifted with respect to the emitted wavelength (and thus also the central wavelength of the direct light line profile). The largest overall blueshift occurs for the scattering element located at point 1 since both the incident and scattered rays gain blueshifts. Conversely, the smallest overall blueshift occurs for point 4, since the scattered ray from this point is redshifted. Line profiles from points 2 and 3 have intermediate blueshifts since only the incident rays gain a blueshift. The overall effect is to broaden and blueshift the integrated scattered line profile relative to the direct light line profile.

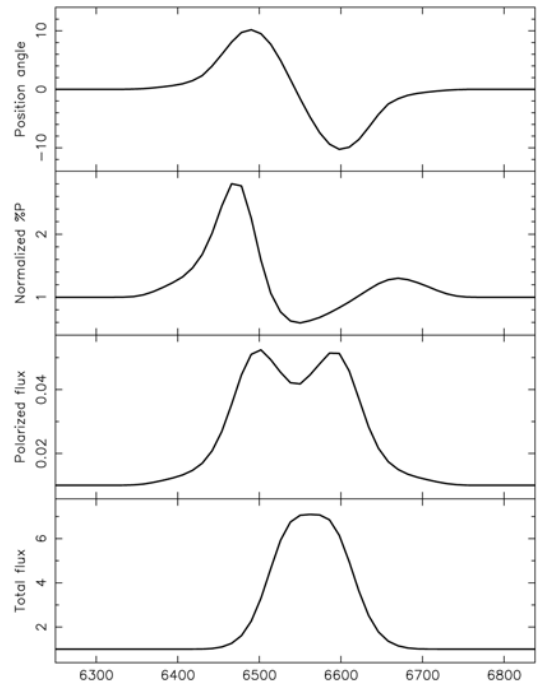

**Figure 10.** As Fig. 4 for a modification of the SEM in which each point in the equatorial scattering region is moving radially inwards at a speed of 900 km s$^{-1}$.

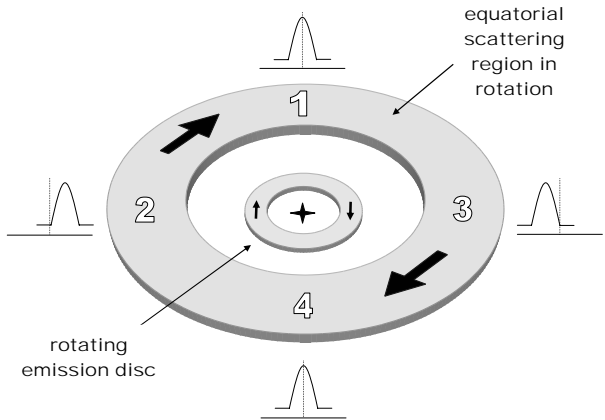

**Figure 11.** Equatorial scattering region undergoing rotation. The doppler-shifts imparted to the scattered line profiles at different points around the region are shown schematically.

Polarization spectra for a model in which the equatorial scattering region is undergoing bulk radial infall with $v_s=900$ km s$^{-1}$ are shown in Fig. 10. These may be compared with the SEM spectra (Fig. 4). The polarized flux line profile is blueshifted relative to the total flux profile and the consequent changes in the relative dilution by unpolarized direct line flux as a function of wavelength results in the strong asymmetry in $p(\lambda)$. The form of the blue-to-red PA rotation is unaffected but its centre is also correspondingly shifted bluewards. A radial *outflow* would produce similar polarization spectra, though reversed in wavelength.

The Seyfert 1's that we identify as showing signatures of equatorial scattering often display an asymmetry in $p(\lambda)$ in the sense that the polarization peak in the blue wing of the H$\alpha$ line is often more prominent than its red wing counterpart. The possibility that such characteristics are explained by bulk inflow of the scattering material is discussed in Section 5.1.

*3.3.2    Rotational motions of equatorial scatterers*

It is natural to suppose that a distribution of equatorial scatterers, co-planar with the rotating emission disc and closely surrounding it, will also be undergoing rotation. Again, for simplicity, we assume uniform motion across the scattering region – i.e., each element has the same rotational velocity $v_s$. In the case of rotational motion, incident rays may be redshifted or blueshifted, depending on the point of origin in the disc. However, the net Doppler-shift for the line profile in incident light is zero. The final doppler-shift of the scattered-flux line profile from a given scattering element is just that corresponding to the component of the scattering element velocity along the observer's line-of-sight, in this case $v_{los}=v_s\sin\alpha\sin i$. The shifts applying to the scattered flux line profiles are shown schematically in Fig. 11 for the points at which the projected major and minor axes cross the scattering annulus. The line profiles from points 2 and 3 gain the greatest doppler-shifts, but in opposite directions [$\pm(v_s/c)\sin i$]. At points 1 and 4 the doppler-shift is zero since the velocity at these points is perpendicular to the line-of-sight. The integrated scattered-flux line profile is broadened as a result of the differential doppler-shifts that are produced around the annulus, but the net wavelength-shift is zero.

In Fig. 12 we present polarization spectra for two models in which the equatorial scattering region is undergoing solid-body rotation. In Fig. 12(a) the scattering region has a rotation velocity of $v_s=1950$ km s$^{-1}$. This is consistent with an extrapolation of the Keplerian velocity law of the emission disc. The rotating scattering region modifies the polarization spectra, as compared with the SEM (Fig. 4), in two main respects. Firstly, the line profile in scattered (polarized) flux is broadened, which has the result that the structure in both $p(\lambda)$ and $\theta(\lambda)$ covers a wider wavelength range. Secondly, a slight inflexion in $\theta(\lambda)$ appears at the central wavelength. If the rotation velocity is increased to $v_s = 2400$ km s$^{-1}$ [Fig. 12(b)] the inflexion develops into a second PA swing in the core of the line profile. This arises because scattered flux from points 2 and 3 in Fig. 11 dominates the resultant polarization and the corresponding line profiles are Doppler-shifted in opposite directions. The elemental scattered line profiles from these points no longer exactly overlap in wavelength, thus separating their corresponding $\theta(\lambda)$ spectra.

While, this is an interesting effect, there is no evidence for such 'double' PA rotations in the data and in any case, it is unlikely that the scattering region will have a higher rotation speed than the local Keplerian velocity. Any such effect is more likely to be on the scale shown in Fig. 12(a) and, therefore, we do not expect rotational motion of the scattering ring to significantly modify the basic

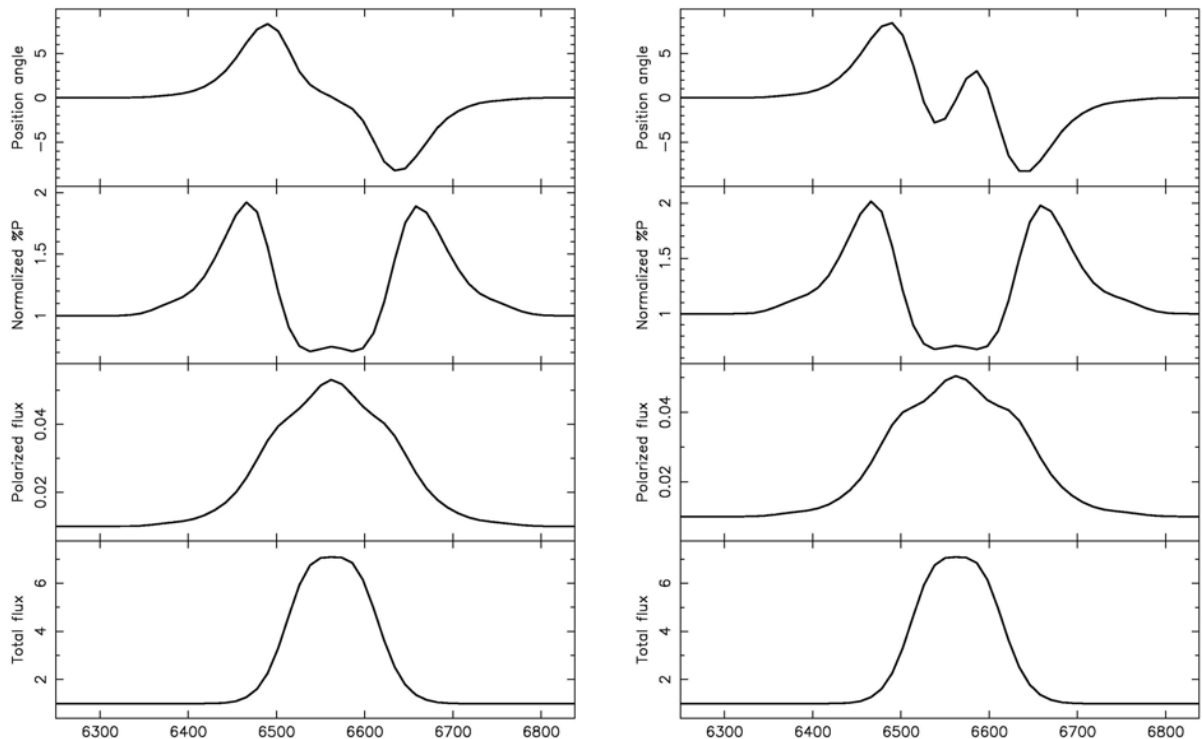

**Figure. 12**. As Fig. 4 for a modification of the SEM in which the equatorial scattering region is undergoing uniform circular motion at speeds of (**a**, left panel) 1950 km s$^{-1}$ and (**b**, right panel) 2400 km s$^{-1}$.

polarization signature of the equatorial scattering model.

## 4  EQUATORIAL AND POLAR SCATTERING

Thus far we have discussed the SEM in isolation. However, we have argued that a polar scattering region is also present in all Seyfert galaxies, including Seyfert 1's (S02; S04). Unlike the equatorial scattering region, the polar scattering cone extends outside the torus, and hence its scattered light is visible in any viewing orientation. The net polarization of a Seyfert 1 nucleus, therefore, results from a combination of equatorial and polar scattering, even in objects where the former dominates. For both components, the monochromatic polarized flux increases monotonically with inclination from zero, at i=0°, to a maximum at i=90°. As shown by S04 (fig. 9), the component that produces the greater polarized flux at any inclination i>0, will dominate at all inclinations (assuming no obscuration by the circum-nuclear torus).

Nevertheless, even when equatorial scattering is dominant, polar scattered light may still have a significant effect on the total polarization spectrum. In particular, light scattered from the polar region is polarized orthogonal to that scattered from the equatorial region. The net degree of polarization at any wavelength is therefore decreased by cancellation between the two polarization states. The resultant polarization PA is also biased towards that of the polar-scattered light, amplifying the PA swing in that direction. Furthermore, polarized broad-lines in Seyfert 2 galaxies tend to exhibit redshifts with respect to the narrow-lines, implying that the polar scatterers are undergoing bulk outflow from the nucleus (e.g. Miller, Goodrich & Matthews 1991; Young et al. 1993). In Seyfert 1's, this would result in cancellation effects between the equatorial and polar-scattered rays occurring asymmetrically across the line profile, wavelengths redward of the line core being most affected.

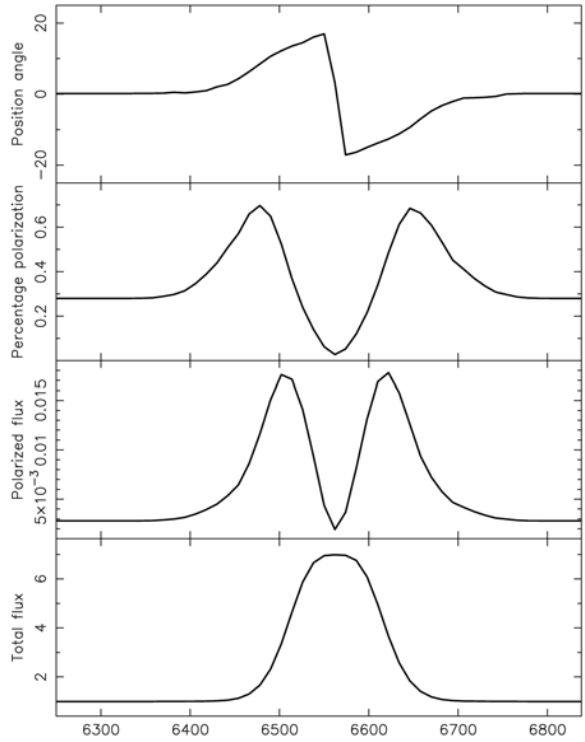

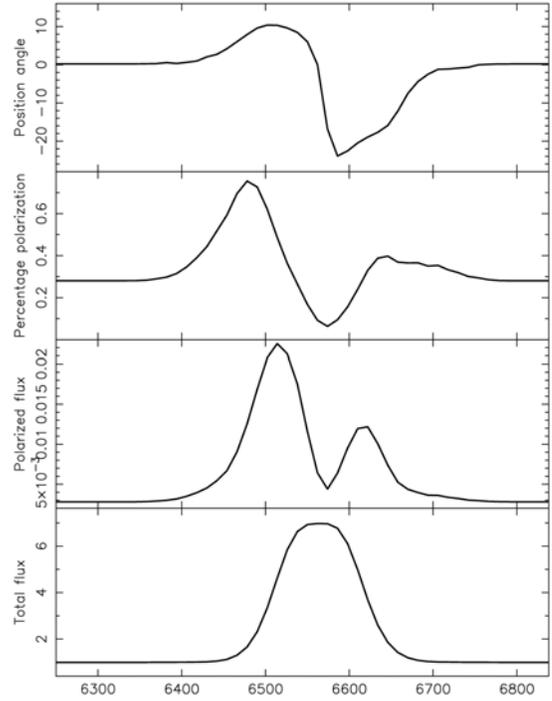

**Figure 13** – continued, and (c) 3000 km s$^{-1}$.

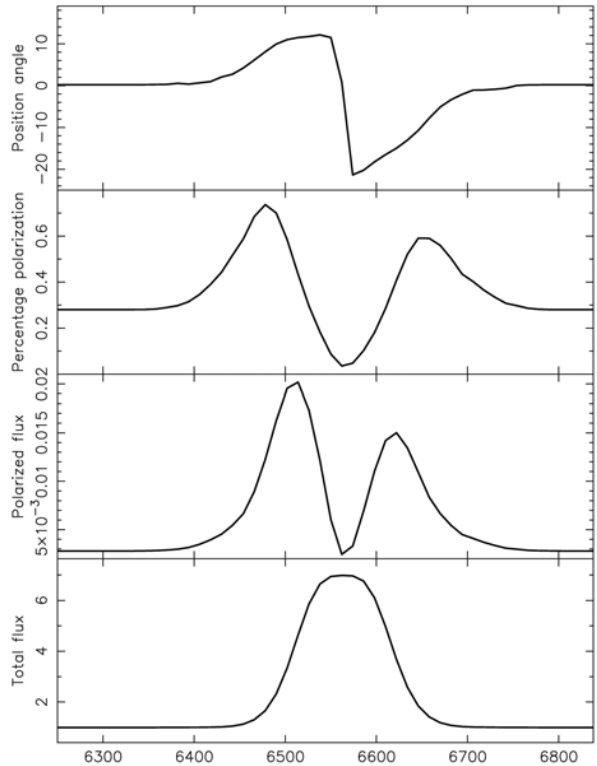

**Figure 13.** The polarization spectra for the case of a two-component polar and equatorial scattering geometry in which the scattering cone has an outflow velocity of (**a**, top panel) 0 km s$^{-1}$, (**b**, bottom panel) 1200 km s$^{-1}$.

To investigate these effects in more detail, we have added a polar scattering cone to the SEM. The geometry of this component is described in Section 2.1 (see also S04, fig. 6). The polarization spectra resulting from the combination of equatorial and polar scattering geometry are shown in Fig. 13. The model parameters (Table 2) were chosen such that the equatorial scattering region dominates the total (integrated over wavelength) polarized flux. The main difference with respect to the SEM (aside from the introduction of the polar scattering region) is that the opening angle of the equatorial scattering region is increased from a nominal 1° to 25°.

Since equatorial scattering dominates, the continuum PA is parallel to the system axis, and the general form of the variations in $p$ and $\theta$ over the broad-line in the polarization spectrum are similar to those produced by the SEM (Fig. 4). Model spectra are presented for 3 cases in which the material in the polar scattering cone is participating in a radial outflow whose speed is, respectively, (a) 0 km s$^{-1}$, (b) 1200 km s$^{-1}$ and (c) 3000 km s$^{-1}$. In the first case [Fig. 13(a)], as the polar scattering medium is stationary, the centres of the polarized flux line profiles produced by the two scattering regions coincide. As a result, the opposite PA swings in the blue and red wings are equally biased towards the respective orthogonal PA, increasing the overall amplitude of the rotation across the line profile (relative to the SEM). The line profile in polar-scattered light is narrower than the double-peaked profile in equatorially-scattered light and peaks in the central dip of the latter. Cancellation between the orthogonal polarization states is therefore most effective in the line core, deepening the central trough in the net percentage polarization and polarized flux spectra.

**Table 2.** Parameters for the two-component scattering model.

| | |
|---|---|
| system inclination | 35° |
| emission disc inner radius | $1.0 \times 10^{14}$ m |
| emission disc outer radius | $1.0 \times 10^{15}$ m |
| rotational velocity at inner radii of emission disc | 7500 km s$^{-1}$ |
| compact scattering region half opening angle | 25° |
| compact scattering region inner radius | $1.5 \times 10^{15}$ m |
| compact scattering region outer radius | $2.4 \times 10^{15}$ m |
| electron number density at inner radius of compact scattering region | $3.0 \times 10^{12}$ m$^{-3}$ |
| number density radial dependence in compact scattering region | $r^{-1}$ |
| polar scattering region half opening angle | 45.0° |
| polar scattering region inner radius | $3.0 \times 10^{16}$ m |
| polar scattering region outer radius | $9.0 \times 10^{16}$ m |
| electron number density at inner radius of polar scattering region | $7.7 \times 10^{10}$ m$^{-3}$ |
| number density radial dependence in polar scattering region | $r^{-2}$ |

In the remaining two cases, where the polar scattering medium is undergoing outflow [Figs 13(b-c)], the polarized line profile from the polar scattering cone is redshifted relative to that produced by the equatorial scattering region. Cancellation effects therefore produce asymmetric net polarization spectra. Note in particular, that for the higher outflow velocity [Fig. 13(c)], the red wing polarization peak is largely suppressed. The PA rotation is also asymmetric in the net polarization spectra. Since the polar scattered light falls primarily in the red wing of the line, the net PA on this side is 'pulled' towards the orthogonal PA (90°) of the polar scattered component, producing a larger swing relative to the continuum PA (180°) than in the blue wing.

These models suggest that the contribution of polar-scattered light may explain some of the diversity in the variations in *p* and *θ* across the broad Hα line that is observed in objects whose overall polarization characteristics are otherwise consistent with equatorial scattering. For example, the presence of a redshifted, polar scattered component could explain why the red wing polarization peak is often less prominent than that in the blue wing. Given our proposed two-component scattering geometry, such effects must operate to some extent. However, the outflow velocities required to achieve complete cancellation of the red wing polarization peak are larger (>1000 km s$^{-1}$) than implied by the redshifted polarized broad-lines observed in Seyfert 2 galaxies. A certain amount of 'fine-tuning' of the parameters that determine the amount of polarized light received from each scattering component (the effective covering factor) is also required, such that polar scattering significantly influences the polarization signature of the equatorial component, but does not dominate it.

The characteristic polarization spectrum produced by the near-field equatorial scattering region might similarly be affected by other sources of scattered, polarized light. For example, scattering from the inner wall of the circum-nuclear torus may contribute to the total polarization (Young et al. 1999). However, since this is also 'equatorial scattering', albeit in the far-field, the polarization PA will be parallel to the symmetry axis, as for the compact scattering region and therefore, does not greatly alter the net polarization spectrum. Another possibility is that isolated clouds of scattering material contribute polarized light with a random PA and at a random doppler shift. However, while the ad hoc introduction of such structures could certainly help to account for some of the observed diversity in $p(\lambda)$ and $\theta(\lambda)$, this is not in keeping with our main aim of formulating a general picture for the scattering geometry.

## 5 DISCUSSION

If the source of the broad emission-lines is a rotating disc and the optical continuum comes either from the disc or a central point source, scattering in an equatorial ring co-planar with the disc and closely surrounding it will, in general, produce polarized light with the following characteristics:

(i) In the underlying continuum, the position angle of polarization is aligned with the projected disc rotation axis and hence also with the radio source axis;

(ii) The polarization PA rotates across the broad emission-line profile, reaching equal but opposite (relative to the continuum PA) rotations in the blue and red wings;

(iii) The degree of polarization peaks in the line wings and passes through a minimum in the line core. The underlying continuum is polarized at an intermediate level.

As discussed in S02, one or more of these distinctive features are commonly observed in our sample of Seyfert 1 galaxies. However, there is also much diversity in detail between objects. Here, we compare the predictions of our equatorial scattering model to the observed polarization spectra and discuss how the modifications to the basic model described in Sections 3 and 4 may explain certain deviations from expected polarization characteristics. We go on to consider the implications of the observed polarization structures for the nature of the BLR and finally, speculate on the nature of the equatorial scattering region itself.

### 5.1 Comparison with observed polarization characteristics

In S02 we identified 10 Seyfert 1 nuclei whose broad H$\alpha$ emission-lines show either opposite rotations in $\theta$ in the blue and red wings (e.g. Mrk 985; fig. 15 of S02) *or* a decrease in the degree of polarization at the line core, flanked by peaks in $p$, above the continuum level, in either line wing (e.g. Mrk 6; Fig. 14). In addition to our own sample, one of the objects studied by Martel (1996; 1998), NGC 4151, also shows clear signatures of equatorial scattering. The polarization properties of these 11 objects are summarized in Table 3.

**Table 3.** Objects exhibiting polarization characteristics consistent with the equatorial scattering model.

| Object | PA rotation across broad H$\alpha$ line profile? | Depolarization in broad H$\alpha$ line core? | Polarization peak(s) in either broad H$\alpha$ line wing? | Continuum polarization PA parallel to radio axis? |
|---|---|---|---|---|
| Akn 120 Epoch 1 | Yes | Yes | No | Yes |
| Epoch 2 | Yes | No | Yes | |
| IZw 1 Epoch 1 | No | Yes | No | Yes |
| Epoch 2 | Yes | Yes | No | |
| KUV 18217+6419 | Yes | No | Yes | No |
| Mrk 6 | Yes | Yes | Yes | Yes |
| Mrk 304 | No | Yes | Yes | No |
| Mrk 509 All Epochs | No | Yes | Yes | Yes |
| Mrk 841 | No | Yes | Yes | ? |
| Mrk 876 | Yes | Yes | No | Yes |
| Mrk 985 | Yes | Yes | No | ? |
| NGC 3783 | No | Yes | Yes | ? |
| NGC 4151 | Yes | Yes | Yes | Yes |

Mrk 6 and NGC 4151 provide the best overall match to the qualitative predictions of the standard equatorial scattering model. The data for Mrk 6 are reproduced here for convenience (Fig. 14). The remaining objects listed in Table 3 exhibit some but not all of the predicted polarization characteristics. Clearly, these features are present to very different degrees in different objects.

It is important to note that the rotation in $\theta$ and the structure in $p$ result from *different* mechanisms (an anisotropy in scattering angle distribution and differential dilution, respectively). Although both effects depend on the line-emitting disc and scattering region being inclined to the line-of-sight, they are also subject to other parameters, which can independently affect their magnitudes. For example, the amplitude of the PA rotation is sensitive to the emission-disc scattering ring distance (Section 3.2), while the percentage polarization structure can be modified by the relative strengths of additional emission or scattering sources (see discussion below). Allowing for object-to-object variations in the structure of the BLR and scattering regions, therefore, the absence of a one-to-one correspondence between the $\theta$ and $p$ structures is not in itself a serious problem for the model.

Another potential source of diversity is polarized flux from the three narrow lines (the narrow component of H$\alpha$ itself and the [NII] $\lambda\lambda$ 6548,6583 doublet) that fall within the broad H$\alpha$ profile. In some objects, the narrow lines exhibit a different polarization state to those of either the continuum or the broad lines. For example, in Mrk 841, the [OIII] $\lambda\lambda$4959,5007 doublet appears to be polarized at PA perpendicular to that of the continuum (S02, fig. 13). Clearly, if the narrow lines are strong enough, cancellation or dilution could affect the polarization structure across broad H$\alpha$. However, close inspection of the spectra (see S02) shows that in most of the objects listed in Table 3, the narrow lines are relatively too weak and too narrow to significantly distort the variations in $p$ and $\theta$ across the broad H$\alpha$ profile. The exception is NGC3783, where the sudden change in $\theta$ in the core of the H$\alpha$ profile is probably largely due to the narrow lines (S02; fig. 17), and as such is not regarded as an equatorial scattering feature. Although Mrk 6 has relatively strong and broad narrow lines, the fact that the [OI] $\lambda$6300 and [SII] $\lambda\lambda$6717,6731 lines do not appear in the polarized flux spectrum (Fig. 14) suggests that they are not strongly polarized in this object.

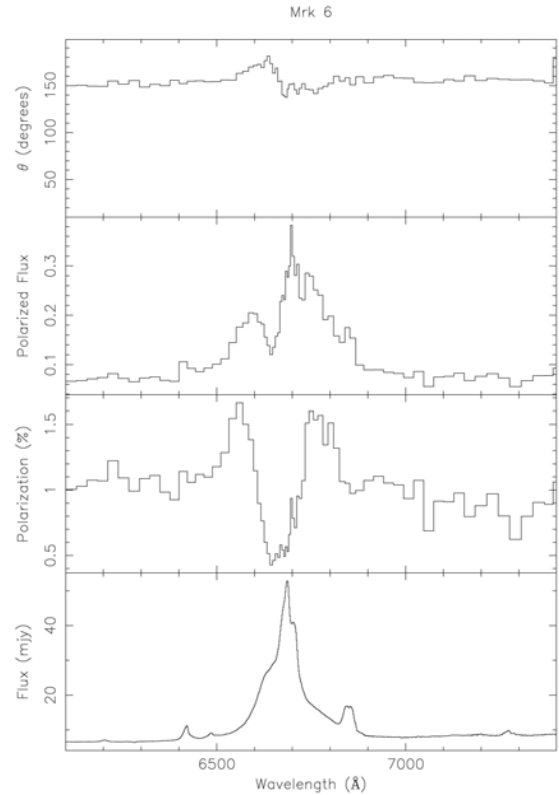

**Figure 14.** Spectropolarimetric data for Mrk 6. The panels show, from the bottom, the total flux density, the percentage polarization, the polarized flux density and the position angle of polarization ( ). Polarization data binned at 0.1 per cent.

### 5.1.1 *Relationship between polarization PA and radio source axis*

An important prediction of the SEM is that the continuum and average line polarization should, in projection, be parallel to the symmetry axis, and hence the radio axis. We have obtained radio source PA's from the literature for 8 of the objects listed in Table 3 (see S04, table 4 for references). Considering the continuum polarization averaged over wavelength, we find that in 6 objects the difference in PA ($\Delta$PA) with respect to the projected radio axis is <30°, consistent with the equatorial scattering interpretation. The two remaining objects are KUV 18217+6419 and Mrk 304. Blundell & Lacy's (1995) 8-Ghz radio map of KUV 18217+6419 shows a strong unresolved core surrounded by various components of extended emission. These authors propose that the jet axis is defined by a pair of relatively faint features extending away from the core in opposite directions. If we adopt this interpretation, the average polarization PA differs by $\approx 60°$ from the jet axis. However, the presence of a prominent hot-spot to the SE suggests an alternative interpretation which would place the jet-axis much closer to the polarization PA. This object also exhibits an extreme PA rotation across the H$\alpha$ profile, which may, as discussed in S02, indicate that orthogonal polarization states (as from equatorial and polar scattering) dominate the blue and red wings.

The case of Mrk 304 has been previously discussed in S02 and S04. It exhibits a clear peak-trough-peak variation in $p$ across the H$\alpha$ profile, but no PA rotation. If we take at face value the radio source PA quoted by Martel (1996; obtained second-hand from a private communication, apparently based on unpublished observations) then $\Delta PA \sim 90°$, consistent with polar rather than equatorial scattering. However, as discussed in S02, partial obscuration of an equatorial scattering region can leave a 'scattering arc' which will produce polarization perpendicular the symmetry axis and hence the radio axis (see also Corbett et al. 1998).

*5.1.2  Modifications to the SEM*

We now consider whether simple modifications of the SEM can account for some or all of the observed diversity in $\theta$ and $p$ over the broad H$\alpha$ profile. The most significant observed deviations from predicted forms $\theta(\lambda)$ and $p(\lambda)$ are:

(i)   the absence of a PA rotation across H$\alpha$ in objects which exhibit the expected variation in $p$ (e.g. Mrk 509; S02, fig. 12);
(ii)  an asymmetry in the PA rotation in the sense that it is not centred on the continuum PA (e.g. Mrk 6);
(iii) a wide variation in the prominence of the percentage polarization peaks in the line wings.

In terms of the model, the simplest explanation for the lack of a PA rotation is that the emission disc – scattering region distance is large (Section 3.2). Adjusting this parameter has little effect on $p(\lambda)$, so it is possible for a given object to exhibit strong variations in $p$, without significant variations in $\theta$. The amplitude of the PA rotation is also sensitive to the surface emissivity distribution of the BLR and the radial density distribution of the scattering medium. Both of these parameters influence the distance between the *effective* (emissivity-weighted) disc radius and the *effective* (density-weighted) radius of the scattering region.

In the SEM the PA rotation over the broad H$\alpha$ line is centred on the continuum PA in the sense that the blue and red wings exhibit swings away from this PA of the same amplitude but in opposite directions. The continuum PA lies within the range covered by the H$\alpha$ PA rotation in 6 out of the 7 objects that exhibit such structure (I Zw 1 being the exception), but in only two (Mrk 876 & Mrk 985) is it approximately centred with respect to the blue and red wing PA's. As discussed in Section 4, asymmetries in both $p$ and $\theta$ can be produced if the polar-scattering region is undergoing bulk radial motion and orthogonally polarized light from this region makes a significant contribution to the net polarization. In particular, the PA rotation across H$\alpha$ becomes asymmetric relative to the continuum because, in the case of outflow, the red wing PA is biased towards the orthogonal PA of the polar-scattered light. However, while this effect must occur to some extent if all Seyferts have both equatorial and polar scattering regions, it will not significantly affect the PA rotation if the polar outflow speeds are comparable with those inferred in Seyfert 2 galaxies.

Another possibility is that the continuum PA is modified by an additional component of polarization that is *not due* to scattering in the compact equatorial scattering region. The similarities between the average continuum and broad H$\alpha$ PA's suggest that the observed continuum polarization is largely due to equatorial scattering 'downstream' of the BLR. However, in the conventional picture of AGN, the optical continuum source (the inner accretion disc) is located interior to the BLR. Continuum emission may, therefore, encounter other scattering material, located between the continuum source and the BLR (e.g. Kishimoto, Antonucci & Blaes 2003) on its escape from the nucleus. Another potential source of 'upstream' continuum polarization is an electron scattering atmosphere above the accretion disc itself (e.g. Coleman & Shields 1990; Laor, Netzer & Piran 1990; Kartje 1995; Antonucci 2001). A component of polarization produced by 'upstream' scattering may, therefore, introduce an offset between the continuum PA and the average line PA.

We now turn to the diversity in the observed percentage polarization structure over the broad H$\alpha$ line. Only in Mrk 6 and Mrk 509 do the observed variations in *p* closely match the model prediction of a central dip in the core of the line profile flanked by peaks of roughly equal height, above the continuum level, in the red and blue wings. More commonly, the polarization peak is stronger in the blue wing than in the red wing (e.g. Mrk 304, Mrk 841 and NGC 3783). In some objects, which do, however, exhibit PA rotations, these features are apparently absent and only the central dip in *p* is observed (I Zw 1, Mrk 876, and Mrk 985). Yet other objects (KUV 18217+6419 and Akn 120 at epoch 2) exhibit large PA rotations and blue wing polarization peaks, but not the central dip in *p* the core of the line.

The asymmetry in the amplitudes of the blue and red wing polarization peaks can be explained by the contribution of polarized flux from the polar scattering region, given a sufficiently high outflow velocity (Section 4). However, as already noted, the required velocity is inconsistent with the available observational evidence from polarized broad-lined Seyfert 2's.

Bulk radial motions in the equatorial scattering region can also produce an asymmetry in the amplitudes of the blue and red wing polarization peaks. Scatterers undergoing radial infall will produce a blue-shifted line profile in scattered light, leading to asymmetric dilution of the resulting polarization by the direct line profile (Section 3.3.1; Fig.10). However, in such a case, any associated PA rotation will also blueshifted (Fig 10), and in both Akn 120 (epoch 2) and KUV 18217+6419, the PA rotation largely occurs in the red wing of the H$\alpha$ line, presenting problems for this interpretation.

Neither of these mechanisms, while capable in principle of producing the observed asymmetry, is entirely satisfactory. Objects which lack polarization peaks in the line wings altogether, or which lack a polarization dip in the line core are even more problematic. There is no natural explanation for such cases within the confines of the pure two-component scattering model.

However, both the BLR and the surrounding scattering regions are undoubtedly much more complex than can easily or usefully be represented in a simple generic model. Direct evidence for complexity in the BLR comes from the broad H$\alpha$ line profiles in total flux, which frequently exhibit asymmetries and other structure that cannot be reproduced by a uniform thin disc in Keplerian rotation. Such structures imply deviations from circular symmetry in the disc (e.g. warping, non-zero ellipticity or 'hot spots') or the presence of an additional component of emitting gas with different geometry and kinematics (c.f. Section 5.3). Unfortunately, our current understanding of the BLR itself does not extend to the detailed shapes of the line profiles, which moreover, differ widely from object-to-object. Incorporating these effects in our model would require many additional free parameters and hence produce little new insight. We note, however, that both asymmetries in the line-emitting disc and the presence of additional components of line-emitting gas will modify the polarization signature of the basic equatorial scattering geometry in ways that are highly model-dependent and hence difficult to predict.

Departures from circular symmetry in the equatorial scattering ring will also affect the polarization across the broad H$\alpha$ profile. One way in which such an asymmetry could arise is through partial obscuration of the scattering ring. In our model, obscuration of the equatorial scattering region by the relatively tenuous atmosphere of the torus explains why some Seyfert 1's are dominated by polar-scattering (S04). It is also possible that only part of the equatorial scattering ring is obscured, producing a 'scattering arc' in the equatorial plane of the torus. Such a configuration will, in general, produce asymmetries across the H$\alpha$ profile in both $\theta(\lambda)$ and $p(\lambda)$ (c.f., Section 2.2.2; see also CM01).

In general, it seems reasonable to conclude that the diversity in the observed polarization properties simply reflects asymmetries and complexity in the emission and scattering regions, which differ in detail between objects. Complex, non-axisymmetric structure is directly observed in the polar scattering region in the archetypal Seyfert 2 galaxy NGC 1068 (Capetti et al. 1995), and it would be naïve not to expect similar complexity in the equatorial scattering region. There is little value in attempting to interpret these detailed differences in the context of our generic scattering model.

## 5.2 Constraints on the scattering geometry from the amplitude of the PA rotation

The amplitude of the PA rotation across the broad H$\alpha$ line is determined by a number of parameters, including the inclination of the symmetry axis and the separation between the emission disc and scattering ring. As discussed in Section 3.1, the PA amplitude is relatively insensitive to inclination over the range $(0 \leq i \leq 45°)$ in which we expect to observe Seyfert 1 nuclei. Other parameters influence the PA amplitude by altering the effective distance between the emission and scattering centres. These are the radial thickness of the emission disc and its radial emissivity distribution and the radial thickness and radial density distribution of the scattering ring. The observed PA amplitude, therefore, is potentially a powerful constraint on the structures of the emission disc and scattering region, as well as their separation.

In the SEM (Fig. 4) the PA rotation amplitude is $\sim 20°$. However, much larger PA rotations are seen in some objects, reaching $\sim 70°$ in continuum-subtracted spectra (e.g. Akn 120; S02, fig. 24), and $\sim 40°$ without continuum subtraction. To reproduce such large PA rotations in the model requires an emission disc – scattering region geometry with the following features:

(i) the separation between the outer edge of the emission disc and the inner edge of the scattering region must be small, so the emission disc subtends a sufficiently large angle at the effective radius of the scattering region;

(ii) scattering must occur predominantly at the inner edge of the scattering region to avoid dilution of the PA rotation by light scattered at larger radii where the emission disc subtends a smaller angle;

(iii) the emission 'disc' itself must be a relatively narrow annulus.

The requirement that the line-emitting disc is actually a relatively narrow annulus has implications for the shape of the line profile. In general, simple Keplerian discs covering a small range in radius produce line profiles with prominent, widely separated double peaks (Robinson 1995). In our model, double-peaked line profiles should be most prominent in polarized flux, since the scatterers have an 'edge-on' view of the disc. The polarized flux spectra have relatively low signal-to-noise ratios and are difficult to characterize accurately. However, there is at least one object in our sample, Mrk 6 (Fig. 14) in which the broad H$\alpha$ profile has a clearly double-peaked structure in polarized flux. There is little evidence for double-peaked profiles in the total flux spectra. While a number of objects have H$\alpha$ profiles that exhibit prominent shoulders on the red or blue side of the line profile (including Mrk 6 itself), none have clearly double-peaked structures.

The equatorial scattering model therefore appears to be subject to two apparently conflicting constraints. In order to obtain the large observed PA rotations, we require that the emitting region is a narrow rotating annulus. Yet such a structure produces double-peaked line profiles that are not observed, at least in total flux. One possible resolution of this problem is to require that the surface emissivity distribution within the disc is a steep function of radius; the double peaked structure becomes less prominent as the slope of the emissivity distribution is increased (Robinson 1995). Another possibility, which we will discuss in more detail below (Section 5.3) is that the BLR has another component, in addition to the rotating disc, and emission from this component 'fills in' the central dip in the double peaked profile produced by the disc.

## 5.3 Implications for the broad-line region

Despite numerous observational and theoretical studies (e.g. Sulentic, Marziani & Dultzin-Hacyan 2000), the structure and dynamics of the BLR remain poorly understood. Although there is no clear consensus, it has frequently been postulated that low ionization broad-lines (LIL) such as H$\alpha$ and H$\beta$ are emitted by a rotating disc, presumably the outer regions of the accretion disc (e.g. Shields 1978; Collin-Souffrin 1987). However, high ionization lines (HIL) such as CIV $\lambda$1550 often exhibit properties more consistent with a radial flow than with orbital motions, particularly in higher luminosity AGN (e.g. Espey et al. 1989; Leighly & Moore 2004). Such considerations have prompted models in which the HIL originate from an outflow or wind along the system axis, whilst the LIL originate from a more flattened disc-like system (e.g. Collin-Souffrin et al. 1988; Marziani et al. 1996;

Murray & Chiang 1997; Gaskell & Snedden 1999; Gaskell 2000).

The most striking signature of disc emission is a double-peaked broad-line profile. Double-peaked Balmer-lines appear to be relatively common in broad-line radio galaxies (Eracleous & Halpern 1994, 2003; and are also found, often as transient features, in some LINERs (e.g. Storchi-Bergmann, Baldwin and Wilson 1993; Bower et al. 1996; Ho et al. 2000; Shields et al. 2000; Storchi-Bergmann et al. 2003). More generally in radio-loud quasars, the width of the broad H$\beta$ line is inversely correlated with various parameters measuring radio core dominance, consistent with a disc origin for the Balmer line emission. (Wills & Browne 1986; Wills & Brotherton 1995).

Double-peaked Balmer-lines are comparatively rare in radio-quiet AGN. Strateva et al. (2003) have recently identified 86 objects with double-peaked broad H$\alpha$ lines in a uniformly-selected sample of AGN taken from the Sloan Digital Sky Survey. The majority of these (75 per cent) are radio-quiet. However, the candidate double-peaked H$\alpha$ emitters represent only 3 per cent of the sample, and only $\approx 1/3$ of these clearly exhibit 2 separate peaks (most identifications are based on model fits to 'blended peaks' or 'shoulders' in the blue or red wings).

The H$\alpha$ polarization of a small number of radio-loud broad-line AGN has been studied by Corbett et al. 1998; Kay 1999 and Corbett et al. 2000. At least one source, 3C445, exhibits a clear PA rotation and evidence for a peak-trough-peak variation in $p$ across the line profile (Corbett 1998; CM01). Curiously, however, the objects that exhibit the clearest double-peaked line profiles do not show these characteristic polarization signatures of equatorial scattering of line emission from a rotating disc. Indeed, Corbett et al. (1998; 2000) have argued that the H$\alpha$ polarization properties of these objects are best understood if the line emission comes from a bi-polar outflow, rather than a rotating disc.

The evidence from the Balmer-line profiles that part of the BLR exists in the form of a line-emitting disc is, therefore, both ambiguous, and at least for Seyferts, limited to a small fraction of the population. Similarly, although reverberation mapping (e.g. Peterson 1993) is potentially capable of mapping the structure of the BLR, this highly demanding technique has thus far failed to provide unambiguous results. Reverberation mapping studies of several Seyfert 1's favour virialized gas motions (e.g. Peterson & Wandel 2000; Onken & Peterson 2002), but do not strongly constrain the precise structure of the BLR. In this context, the distinctive variations in $p$ and $\theta$ across the broad H$\alpha$ line that we have found to be relatively common in Seyfert 1 galaxies are of particular interest. Although it is possible to conceive of other mechanisms that may explain either the PA swing or the structure in $p$ across the line profile, the only plausible emission source/scattering region geometry that naturally accounts for both of these properties is a rotating emission disc surrounded by a co-planar scattering ring. Arguably, therefore, our broad H$\alpha$ polarization data represents the most compelling evidence yet that a significant fraction of the broad-line emission does indeed originate in a disc.

However, if this is correct, the general question of why double-peaked line profiles are so rarely observed remains to be explained. As already noted in Section 5.2, even those objects of our sample that exhibit clear signatures of equatorial scattering in their polarization spectra do not, with the exception of Mrk 6 in polarized flux, exhibit clearly double-peaked line profiles. A possible explanation is that the BLR contains two components of Balmer line emitting gas — a rotating disc, as required to produce the polarization properties, and a second component producing a centrally-peaked intrinsic line profile. There is, indeed, evidence for a central core component in the Balmer lines of broad-line radio galaxies, including those which exhibit clearly double-peaked profiles (Halpern & Eracleus 1994; Halpern et al. 1996).

One possibility, which would be consistent with reverberation mapping results, is that the Balmer lines have a central peak produced by a spherical ensemble of clouds following virialized orbits. The equatorial scattering region will scatter H$\alpha$ emission from both components, but only for the disc component will scattering produce wavelength-dependent polarization. Since the spherical cloud ensemble has an isotropic velocity field, there is no spatial discrimination, on average, between red and blueshifted rays and hence no PA rotation over the line. In addition, the line profile will have the same shape and width in scattered and direct flux (e.g. see Fig. 15).

We expect, therefore, that scattering of H$\alpha$ emission from the virialized cloud ensemble will produce constant polarization over the line (i.e. the Stokes parameters $Q$ and $U$ are independent of wavelength). Since the Stokes parameters are additive, the net polarization produced by scattering of H$\alpha$ emission from both the virialized cloud ensemble and the rotating disc will retain the characteristic wavelength dependence of the latter (although the wavelength-averaged values of $p$ and $\theta$ will be modified).

Therefore, even though the cloud ensemble might contribute significantly to the total flux spectrum, the variations in $p$ and $\theta$ across the broad Balmer lines are still largely determined by scattering of the disc component. On the other hand, the double-peaked line profile produced by the disc could easily be masked (in both total and polarized flux) by low velocity emission from the virialized cloud ensemble.

An alternative scenario is that the BLR consists of a rotating disc and an outflowing wind. As already noted, this type of structure is postulated in several models (e.g. Murray & Chiang 1997). The main observational evidence for a radial outflow comes from blueshifts and asymmetries in HIL profiles (e.g. Espey et al. 1989), implying a high ionization state. However, the wind will also be a source of Hydrogen recombination emission, which will contribute to the Balmer lines, the bulk of which are presumed to come from the disc. It is, therefore, of interest to briefly consider the polarization signature that would be produced by a bi-polar radial outflow in our two-component scattering geometry.

If the outflow axis is aligned with the system symmetry axis, the equatorial scattering region will tend to see a narrower line profile than the direct line-of-sight. In general, the light incident on the scattering region will also have a net redshift since the bulk of the flow is receding from the scattering region. On the other hand, the directly observed line profile may have a net blueshift if part of the receding (with respect to the observer) flow is obscured, for example by the accretion disc (e.g. Livio & Xu 1997). Thus, the line profiles in polarized and direct flux are likely to have different asymmetries and widths which will in turn produce asymmetric structure in $p$ across the line — for example, a dip in the blue wing and a peak in the red wing. In general, we expect that the polarization PA will be constant across the profile and parallel to the axis but more complex behaviour is also possible, depending on the geometry and velocity field of the flow. We will explore the emission-line polarization produced by a bi-polar radial outflow within our two-component scattering geometry in more detail elsewhere. Here we simply note that if such a flow produces significant Balmer line emission it will contribute to both the direct and scattered flux and thus modify the total and polarized spectra of the disc component. This may help to explain both the lack of clearly double-peaked line profiles (by filling in the central dip) and some of the observed diversity in the polarization structure across the Hα line. The asymmetric Stokes spectra produced by equatorial scattering of emission from the flow will produce asymmetries in the net polarization spectra when added to the disc component (in much the same way as does polarized flux from an outflowing polar-scattering region, Section 4).

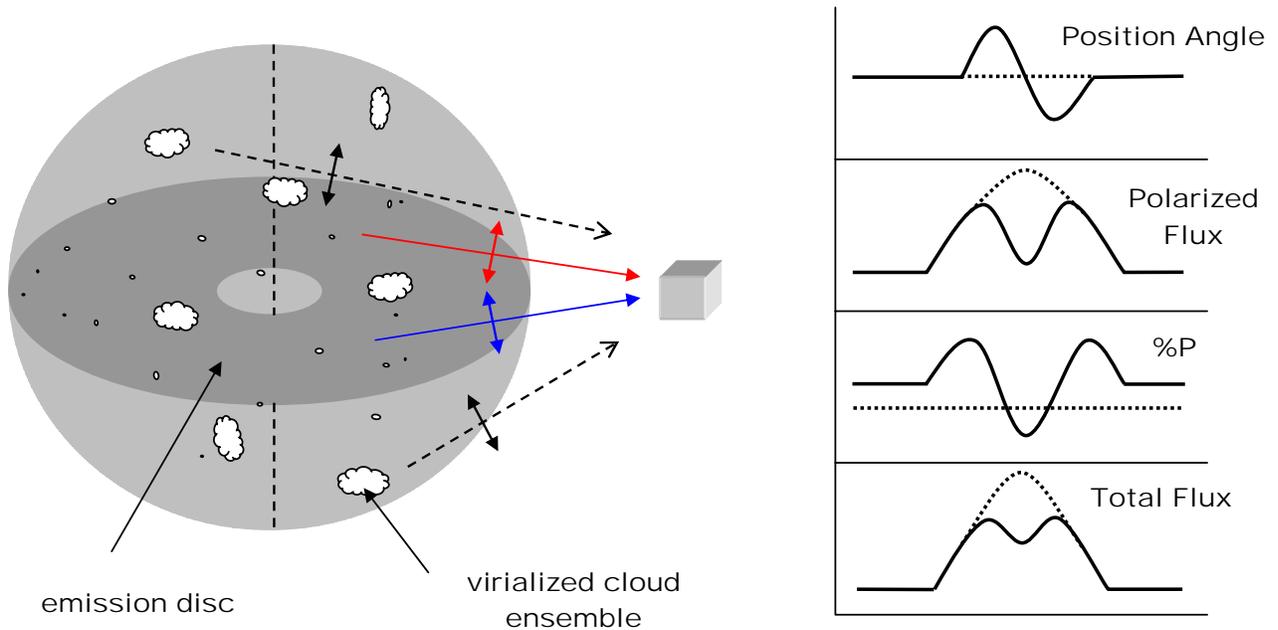

**Figure 15.** The addition of a virialized BLR component to the basic equatorial scattering model. Additional scattering geometries (dashed lines) are introduced whose integrated total flux and polarization spectra add to that produced by the equatorial scattering geometry. In the polarization spectra the components introduced by the virialized BLR component are superposed (dashed line) on the characteristic spectra produced by equatorial scattering of the emission disc (solid line).

In disc/wind models for the BLR, HIL's such as C IV λ1550 are expected to arise only in the high ionization outflow. These lines should therefore exhibit polarization properties quite distinct from those of LIL's, such as Hα or Mg II λ2800, which are emitted predominantly by the disc. Within the range of inclination in which equatorial, rather than polar, scattering is expected to dominate (S04), the different emission geometries of LIL's and HIL's should lead to different polarization signatures and therefore, high quality spectropolarimetry of representative HIL and LIL may prove a fruitful way of investigating the structure of the BLR and of testing disc/wind models in particular. Unfortunately, suitable HIL's are only found in the rest-frame UV, so with present facilities, such an investigation can only be applied to bright quasars at redshifts $z \sim 2$, sufficient to shift these lines into the optical regime.

To conclude this section, equatorial scattering of line emission originating from a rotating disc provides a natural explanation for the two most striking features of the Balmer line polarization, namely the peak-trough-peak variation in $p$ and the double rotation in $\theta$. However, in detail, the observed line profiles and differences between objects in $p(\lambda)$ and $\theta(\lambda)$ cannot be completely explained in terms of equatorial scattering of light from a simple rotating disc alone. Rather, it seems likely that there is another source of Balmer line emission present in the BLR, which contributes to both the directly-viewed and scattered fluxes.

### 5.4 Nature of the equatorial scattering region

We first invoked a 'compact' scattering region in order to explain temporal variations in the polarized light of Mrk 509 (Young et al. 1999). The alignment of the polarization PA with the radio axis in this object also led to the inference that this scattering region lies in the equatorial plane of the system. In Young et al. (1999) we speculated that it could be associated with 'the optically thin atmosphere of the accretion disc, or possibly the region connecting the torus to the accretion disc, or in the base of the outflow which produces the radio jet'. The need to explain the PA rotation across the broad Hα line leads to the further refinement that the equatorial scattering region is co-planar with and closely surrounds the rotating disc that produces the bulk of the Balmer line emission.

The observations do not, however, strongly constrain the vertical height of the scattering region. In our model, the polarization spectrum is relatively insensitive to the opening angle of the equatorial scattering region, as long as this does not greatly exceed 45°. Indeed, a significant scale-height is essential if the equatorial scattering is to produce enough polarized flux to dominate the polarization spectrum in most Seyfert 1's (S04). In hydrostatic equilibrium, the scale height-to-radius ratio of the scattering region is $\sim c_S/v_S$, the ratio of the sound speed to the Keplerian velocity. This ratio is on the order of unity for a free electron gas with plausible values for the temperature ($3 \times 10^5$ K; see below) and Keplerian velocity (1950 km s$^{-1}$; Section 3.3.2). Therefore, it is entirely plausible that the scattering region has a scale height comparable to its radius. Its geometry may be more similar to a torus than to the thin wedge adopted (mainly for simplicity) in the model.

We have assumed in our model and in the preceding discussion that the equatorial scattering region is composed of free electrons. Although we lack the wavelength coverage to distinguish observationally between dust and electrons, this assumption is justified by the location of the scattering region: within the circum-nuclear torus and closely surrounding the BLR. This places it well within the dust sublimation radius, which presumably coincides with the inner edge of the torus. The complex polarization variations seen over the broad Hα line imply that the electrons are relatively cool. If the electron temperature greatly exceeds a few $\times 10^5$ K thermal Doppler broadening will effectively smear out this structure (Smith 2002).

Intriguingly this upper limit lies within the temperature range inferred for the highly ionized gas responsible for various absorption features often observed in the X-ray spectra of many Seyfert galaxies — the so-called 'warm absorber' (e.g. Turner et al. 1993). Warm absorbers have been detected in several of the objects in our sample, including some of the objects identified as being dominated by equatorial scattering (e.g. Mrk 509, Perola et al. 2000; NGC 3783, George, Turner, & Netzer 1995; NGC 4151, Schurch & Warwick 2002). The location and origin of this gas is unclear. It has been variously argued that the warm absorber is associated with the BLR (George et al. 1998), or is a wind driven off the accretion disc (Elvis 2000; Bottorff, Korista & Shlosman 2000), or is evaporated from the torus (Krolik & Kriss 2001) or is spread out between the BLR and the torus (Morales, Fabian, & Reynolds 2000).

While it is tempting to associate the equatorial scattering region with the warm absorber, geometry dictates that the two occupy different polar angle regimes. The warm absorber must lie in the direct line-of-sight to the central X-ray source and must therefore be located at relatively small polar angles in comparison to the compact scattering region, which occupies the equatorial plane (note that although the scattering electrons may extend to a significant distance above the plane, the absorbing ions will have much smaller hydrostatic scale heights). We note also that scenarios in which a significant column density of warm absorber gas is located above and in close proximity to the BLR are difficult to reconcile with the observed optical polarization. Electron scattering in this gas would be expected to polarize both the broad line emission and the optical continuum, but it is hard to see how the observed large variations in $p$ and $\theta$ over the Balmer lines could be produced in this geometry. Moreover, the polarized light would have an **E** vector perpendicular to the system axis. Leighly et al. (1997), find that Seyfert 1 and 1.5 galaxies which have high broadband optical polarization also tend to exhibit X-ray warm absorbers, suggesting a possible link between the absorbing gas and the polarization mechanism. These authors argue, on the assumption that the polarization is produced by dichroic absorption, that this is evidence that dust is associated with the ionized gas (i.e., a 'dusty' warm absorber). However, dichroism cannot account for the structures in $p$ and $\theta$ associated with the broad Balmer lines that are often observed in Seyfert 1's (S02), which as discussed here, are best explained in terms of equatorial scattering. The relationship between optical polarization and X-ray absorption properties of Seyferts is clearly an area ripe for further investigation.

Note that the preceding arguments do not preclude the existence of a tenuous, much hotter atmosphere. While this will also scatter light from the BLR, for temperatures $>10^7$ K, thermal broadening will tend smear out any wavelength dependence on the scale of the broad lines, leaving a constant pseudo continuum in the Stokes fluxes.

Kishimoto et al. (2003) have recently studied the polarization of the big blue bump continuum emission in two quasars. If the observed polarization is due to external scattering (rather than intrinsic to the big blue bump emitter) the scattering region must be oblate and perpendicular to the system axis. It is geometrically similar to our equatorial scattering region, suggesting that the two may be related. However, the absence of broad lines in polarized flux suggests that the continuum scattering region is located interior to, or possibly co-spatial with, the BLR, rather than exterior to it.

Although the AGN fuelling mechanism is unknown, it seems likely that the accretion disc is replenished by gas infalling from the torus, which represents an immense reservoir of 'fuel'. This being so, mass transfer must take place in the equatorial plane of the torus, and it is tempting to speculate that the equatorial scattering region is associated with this gas flow. In this respect, it is interesting that one possible explanation for the relative prominence of the blue wing polarization peak in H$\alpha$ is that the equatorial scattering region is infalling towards the nucleus. While, at the present stage of our investigation, this can only be regarded at best as very tentative evidence for infall, it does suggest that a combination of high accuracy polarization measurements and detailed modelling may ultimately provide a route to inferring the mass accretion rate in AGN.

## 6 CONCLUSIONS

In this and three previous papers (Young et al 1999; S02; S04) we have endeavoured to set out a general framework, in the broader context of the unification scheme, within which the polarization properties of Seyfert galaxies can be understood. In the general scattering model that we have developed, light from the optical continuum source and the broad-line region is scattered in two separate regions — a polar-scattering region outside the circum-nuclear torus but aligned with its axis and a compact region located within the torus and residing in its equatorial plane. The relative contribution of each scattering region to the net (observed) polarization is determined largely by the inclination of the system axis to the line-of-sight.

When the inclination is near zero (pole-on), cancellation ensures null, or weak, polarization. At intermediate inclinations, where there is no extinction along the direct line-of-sight to the nucleus, both scattering regions, as well as the broad-line region, are visible but in general equatorial scattering dominates the observed polarization. When the inclination is comparable to the torus opening angle, the direct line-of-sight passes through the upper atmosphere of the torus, polarized light from the compact equatorial scattering region is attenuated and the observed polarization is dominated by polar scattering. At still larger inclinations, both the BLR and equatorial scattering region are completely obscured by the torus and the broad-lines are only visible in polarized light scattered from the polar scattering region (the total flux spectrum is that of a Seyfert 2).

In this paper, we have focussed on the polarization signature of equatorial scattering. We have identified 11 Seyfert 1's that exhibit polarization properties consistent with equatorial scattering, 10 of which are drawn from our S02 sample, representing ~50 per cent of the intrinsically polarized sources. These objects display striking variations in *p* and/or *θ* across the broad Hα line, which are naturally explained if the source of the Hα emission is a rotating disc. The detailed calculations presented here confirm that equatorial scattering of line emission from a rotating disc produces polarization characteristics consistent with those observed. Broad-line emission from the disc has a distinctive polarization signature with the following features:

(i)   averaged over wavelength, the polarization PA is aligned with the projected disc rotation axis (and hence is parallel to the radio source axis);

(ii)   the polarization PA rotates across the broad emission-line profile, reaching equal but opposite (relative to the continuum PA) rotations in the blue and red wings;

(iii)   the degree of polarization peaks in the line wings and passes through a minimum in the line core.

We have also explored the dependence of the polarization spectra on key geometrical parameters of the system. We find that, in order to reproduce the large amplitude of the PA variations seen in some objects, the scattering region must closely surround the line-emitting disc, and the 'disc' must be a relatively narrow annulus. This Balmer-line-emitting annulus can be plausibly identified as a cool outer zone of a larger accretion disc.

The fact that the broad Hα lines in both total flux and (with one exception) polarized flux do not display double-peaked profiles suggests that the BLR contains a second source of Balmer line emission, in addition to the disc. One possibility is a virialized cloud ensemble with spherical symmetry, which could 'mask' the disc profile, without significantly modifying the polarization spectrum. Another, perhaps more interesting, is a bi-polar wind. Equatorial scattering of line emission from a bi-polar outflow is expected to produce a radically different polarization signature from that of a rotating disc.

The observed polarization dictates the basic geometry and general location of the equatorial scattering region. It must be co-planar with, and closely surround, the line-emitting disc and its vertical height to radius ratio must be $\leq 1$. Its location within the circum-nuclear torus suggests that the scattering particles are free electrons. If so, the electron temperature cannot greatly exceed a few $\times 10^5$ K. This limit is within the range of temperatures inferred for the X-ray 'warm absorber', but the geometry of the equatorial scattering region argues against the possibility that the X-ray absorption and optical scattering take place in the same gas column. The asymmetry in $p(\lambda)$ that is often observed across broad Hα may indicate that the equatorial scattering region is undergoing inward radial motion. However, other explanations are equally plausible. For example, contamination by orthogonally polarized light from the polar scattering region will cause a similar asymmetry if the scattering material is moving radially outwards at a sufficiently high speed.

Perhaps our most important observational result is that large variations in both $p$ and $\theta$ across the broad H$\alpha$ line are relatively common. These structures provide arguably the most compelling observational evidence yet that a significant fraction of the broad Balmer-line emission in Seyfert 1 nuclei originates in a rotating disc. An important general implication of the observed PA rotations in particular, is that *the BLR is resolved at the equatorial scattering region*. This being so, we can use spectropolarimetry to obtain unique information on the structure and dynamics of the BLR. Our work shows that line-emitting discs can be identified from their distinctive polarization signature, even when their characteristic spectroscopic signature — a double-peaked line profile — is masked, presumably by emission from other sources. Similarly, spectropolarimetry of high ionization lines may settle the question of whether these arise in a separate component of the BLR, such as a wind. The properties of the scattering region itself are also of great interest since it may be associated with mass transfer from the torus to the accretion disc and hence offers a potential route to estimating the mass accretion rate.

Twenty years ago, spectropolarimetry of Seyfert 2 galaxies played a key role in establishing the unification scheme by revealing, thanks to the existence of the polar scattering 'mirror', the 'hidden' nuclei in Seyfert 2 galaxies. Now, armed with our improved knowledge of the scattering geometry, perhaps spectropolarimetry of Seyfert 1's will have a similarly dramatic impact on our understanding of the broad-line region and the central engine.

## ACKNOWLEDGEMENTS

JES acknowledges financial support from PPARC via a post-graduate studentship and as a post-doctoral research assistant. This research has made use of the NASA/IPAC Extragalactic Database (NED) which is operated by the Jet Propulsion Laboratory, California Institute of Technology, under contract with the National Aeronautics and Space Administration. This work has partly been carried out using facilities and software provided by the STARLINK project.